\documentclass{aa}
\usepackage{graphicx}
\usepackage{txfonts}
\usepackage{mathrsfs}

\begin{document}
\title{Modelling of integrated-light spectra from the optical to the near-infrared:
  the globular cluster G280 in M31}

\author{
    S. S. Larsen
   \inst{1}
   \and
   G. Pugliese
   \inst{2}
   \and
   J.\ P. Brodie
   \inst{3}
}
\institute{
  Department of Astrophysics/IMAPP,
              Radboud University, PO Box 9010, 6500 GL Nijmegen, The Netherlands\\
              \email{S.Larsen@astro.ru.nl}
  \and
  Astronomical Institute Anton Pannekoek, Universiteit van Amsterdam, Postbus 94249, 1090 GE Amsterdam, The Netherlands
  \and
  UCO / Lick Observatory, 1156 High Street,
  University of California, Santa Cruz, CA 95064, USA
}

\offprints{S.\ S.\ Larsen, \email{S.Larsen@astro.ru.nl}}

\date{Received February 5, 2018; accepted May 15, 2018}

\abstract
{In previous papers, we introduced our method for measuring chemical abundances from integrated-light spectra of globular clusters and applied it to a variety of extragalactic star clusters. Our work so far, however, has concentrated primarily on the optical range 4200~\AA -- 6200~\AA.
}
{Here we extend our analysis technique to the infrared and test it on an $H$-band spectrum of the massive globular cluster G280 in M31. 
}
{We simultaneously analyse an optical spectrum of G280, obtained with the HIRES spectrograph on the Keck~I telescope, and an $H$-band spectrum obtained with NIRSPEC on Keck~II. We discuss the sensitivity of our results to various modifications of the input assumptions, such as different line lists and isochrones and the possible presence of a metallicity spread in G280.
}
{When using the most recent version of the Kurucz line list, we measure iron abundances of $\mathrm{[Fe/H]}=-0.68\pm0.02$ from the optical spectrum and $\mathrm{[Fe/H]}=-0.60\pm0.07$ from the infrared spectrum. These values agree well with previous spectroscopic determinations of the metallicity of G280. While the small difference between the optical and infrared measurements is insignificant given the uncertainties, it is also consistent with a metallicity spread similar to that observed in massive GCs such as $\omega$ Cen and G1, and also hinted at by the colour-magnitude diagram of G280. The optical and infrared spectra both indicate an $\alpha$-enhancement of about 0.3--0.4 dex relative to solar-scaled abundances, as typically also observed in Milky Way GCs. 
}
{From this analysis, it appears that our integrated-light analysis technique also performs well in the $H$-band. However,  complications due to the presence of molecular bands and telluric contamination are more severe in the infrared, and accurate modelling of the coolest giants is more critical.
}


\titlerunning{The globular cluster G280 in M31}
\maketitle

\section{Introduction}

Over the past decade, the potential of integrated-light spectroscopy for detailed chemical abundance analysis has been demonstrated by several studies for both old globular clusters \citep{McWilliam2008,Colucci2009,Colucci2017,Sakari2013,Sakari2016,Larsen2012a,Larsen2014,Larsen2017,Larsen2018,Hernandez2018} and young massive clusters \citep{Larsen2006b,Larsen2008a,Gazak2014,Lardo2015b,Hernandez2017,Hernandez2017a}. Compared to spectroscopy of individual stars, analysis of cluster spectra must account for contributions to the integrated light from stars across the Hertzsprung-Russell diagram (HRD), and for a generally non-negligible broadening of spectral features by the internal velocity dispersions of the clusters (typically 5--10 km~s$^{-1}$). 
In very young clusters, with ages of 10--20 Myr, the near-infrared spectra are strongly dominated by the light of red supergiants (RSGs), which simplifies the analysis by allowing the spectra to be modelled as a single equivalent RSG \citep{Larsen2006b,Larsen2008a,Davies2010,Gazak2014,Lardo2015b}. However, in general a full population synthesis is required. 
For old globular clusters (GCs), no single stellar type dominates the light at any wavelength, and work on integrated-light abundance analysis of old GCs has so far concentrated primarily on the optical range.
From observations of Milky Way GCs, it has been found that integrated-light measurements can reproduce abundances measured from individual stars with an accuracy of typically $\sim0.1$ dex \citep{McWilliam2008,Sakari2013,Sakari2014,Colucci2017,Larsen2017}.

While the optical spectra of old stellar populations in general, and GCs in particular, are relatively well understood, the infrared spectral range remains less well explored. \citet{Davidge1990a} obtained $H$- and $K$-band near-infrared spectra of four GCs in M31 and found that they exhibited strong CN absorption bands, as also seen in optical spectra \citep{Burstein1984a,Schiavon2013}. \citet{Lyubenova2010,Lyubenova2012} obtained $JHK$ integrated-light spectra of clusters in the Large Magellanic Cloud and found that a variety of features were well reproduced by simple stellar population models \citep{Maraston2005} for the old GCs in their sample. The spectra of intermediate-age clusters (1--2 Gyr) were less well reproduced, possibly because of difficulties with the modelling of carbon-rich stars. In a recent study, \citet{Sakari2016} measured abundances for many individual elements from $H$-band spectra of a sample of 25 GCs in M31 and found generally good agreement with measurements based on optical spectra from \citet{Colucci2009,Colucci2014}. 

In this paper we carry out a simultaneous analysis of optical (Keck/HIRES) and $H$-band (Keck/NIRSPEC) high-dispersion spectra of the massive M31 globular cluster B225-G280 \citep{Galleti2004}. For brevity, we henceforth refer to the cluster as G280. Applying a single analysis technique over such a broad wavelength range, from the optical to the near-infrared, constitutes a useful consistency check of our approach. With a velocity dispersion of $\sigma=26$~km~s$^{-1}$ \citep{Djorgovski1997} and a corresponding dynamical mass of about $3\times10^6 M_\odot$ \citep{Strader2011}, G280 is one of the most massive GCs in M31, and the cluster has been included in a number of previous spectroscopic studies which found it to be relatively metal-rich. From MMT/Hectospec data, \citet{Caldwell2011} found a metallicity of $\mathrm{[Fe/H]} = -0.5\pm0.1$ and an age of 10.7 Gyr. From optical high-resolution integrated-light spectroscopy, \citet{Colucci2014} measured $\mathrm{[\ion{Fe}{i}/H]} = -0.66$ and $\mathrm{[\ion{Fe}{ii}/H]} = -0.61$ and $\mathrm{[Mg/Fe]} = +0.45$, and estimated the age to be $10\pm3$ Gyr. The analysis of $H$-band APOGEE spectra by \citet{Sakari2016} yielded $\mathrm{[Fe/H]} = -0.64 \pm 0.05$ and abundances of several individual elements.

While the spectroscopic studies agree relatively well on the metallicity of G280, an interesting question is whether a metallicity spread is present within the cluster. Given that a number of the more massive GCs in the Milky Way exhibit significant intrinsic metallicity spreads \citep{Willman2012,DaCosta2015}, it would not be surprising if a considerable spread were also present in G280. With \emph{Hubble Space Telescope} (HST) imaging, it is possible to resolve individual red giant branch (RGB) stars in M31 GCs, and from resolved photometry with NICMOS, \citet{Stephens2001} estimated that the metallicity spread in G280 is relatively small, $\sigma_\mathrm{[Fe/H]} \la 0.2$ dex. They also derived a mean metallicity of $\langle \mathrm{[Fe/H]} \rangle = -0.15\pm0.37$, which is a higher value than is typically found from the spectroscopic studies, albeit with a relatively large uncertainty. From photometry obtained with the Advanced Camera for Surveys (ACS) on board HST, \citet[][F08]{Fuentes-Carrera2008} instead found a much larger metallicity spread of $\sigma_\mathrm{[Fe/H]} = 1.03\pm0.26$ dex. An interesting question is thus whether integrated-light observations can provide additional constraints on the presence of a metallicity spread.

\section{Data}

\subsection{Spectroscopy}

The $H$-band spectra used in this paper were obtained with the NIRSPEC spectrograph \citep{McLean1998} on the Keck~II telescope on Nov~8, 2006. The observations consisted of eight exposures, each with an integration time of 300 s and using an ABBA nodding sequence. We used a $0\farcs432 \times 24\arcsec$ slit. 
The NIRSPEC echellogram contains seven echelle orders with central wavelengths between 1.55~$\mu$m and 1.77~$\mu$m, although the NIRSPEC detector is too small to cover the full echellogram so that gaps are present in the wavelength coverage between the orders. The spectra were extracted with the \texttt{Redspec} IDL package written by L.\ Prato, S.\ S.\ Kim, \& I.\ S.\ McLean, modified to use optimum weighting for the extraction \citep{Horne1986}. The individual one-dimensional spectra were then co-added.
The set-up and data reduction were identical to those described in \citet{Larsen2008a} and we refer to that paper for further details.
The median signal-to-noise per \AA\ for the final reduced spectrum, as estimated from the variance of the co-added pixels at each wavelength bin, was 141 (1.55 $\mu$m), 124 (1.58 $\mu$m), 131 (1.62 $\mu$m), 107 (1.65 $\mu$m), 83 (1.69 $\mu$m), 65 (1.73 $\mu$m) and 42 (1.77 $\mu$m).

The optical spectrum was obtained with the HIRES spectrograph \citep{Vogt1994} on the Keck~I telescope on Nov~1, 1997, using a slit width of $0\farcs9$. It consists of a single 1800 s exposure, which covers the spectral range from 3650~\AA -- 6050~\AA . The HIRES data were reduced with the MAKEE package\footnote{Available at http://www.astro.caltech.edu/\~{}tb/makee/} and the signal-to-noise per \AA, as estimated by MAKEE, is about 110 at 4500~\AA, 160 at 5000~\AA\, and 195 at 6000~\AA .

\subsection{Photometry}
\label{sec:phot}

   \begin{figure}
   \centering
   \includegraphics[width=\columnwidth]{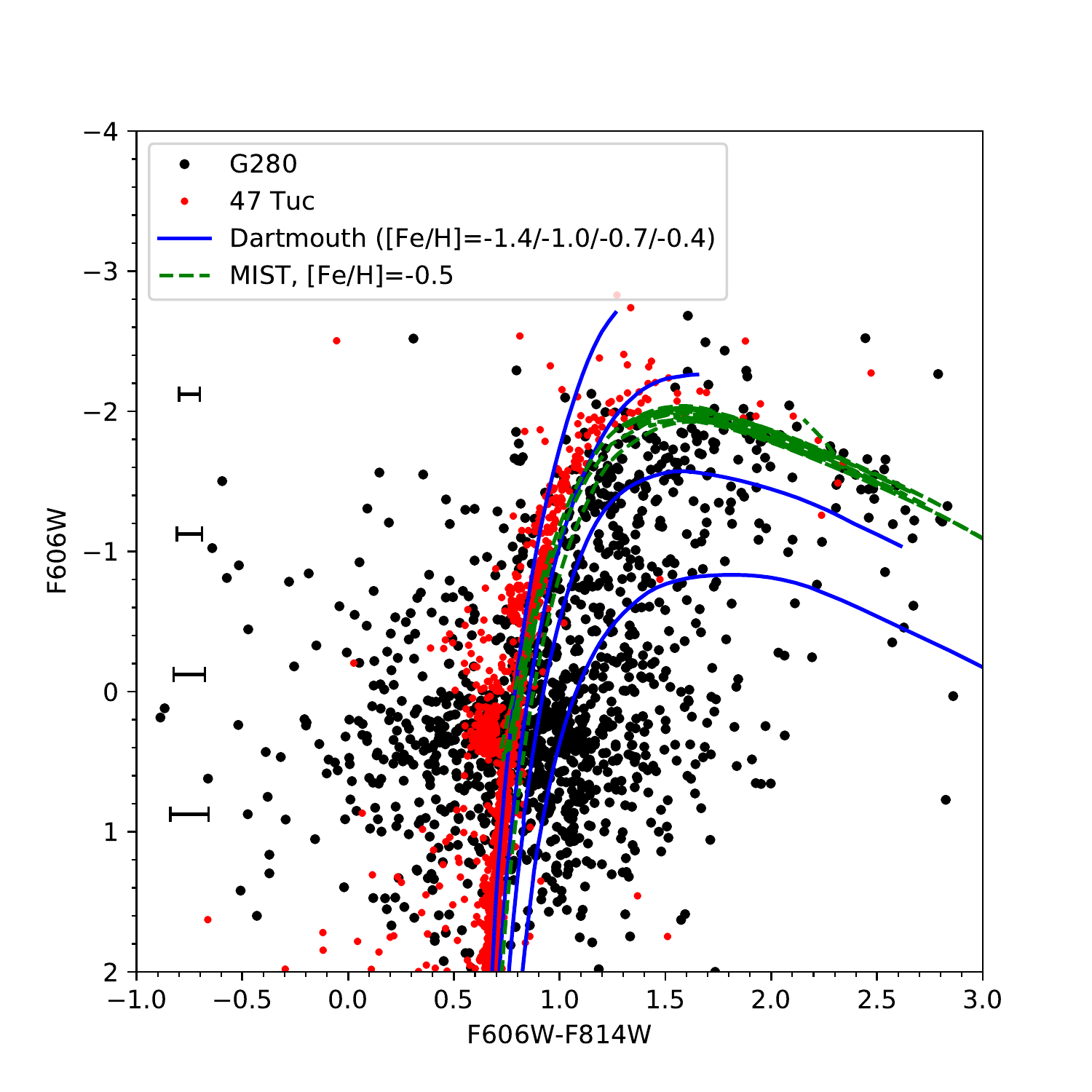}
      \caption{ACS/HRC colour-magnitude diagram for G280. The blue lines are Dartmouth isochrones with ages of 11 Gyr and $\mathrm{[Fe/H]}=-1.4$, $-1.0$, $-0.7$, and $-0.4$ \citep{Dotter2007}, while the green line is a 10 Gyr MIST isochrone with $\mathrm{[Fe/H]}=-0.5$ \citep{Choi2016,Dotter2016}.
      Red dots are ACSGCS data for 47~Tuc \citep{Sarajedini2007}.
         \label{fig:cmd}
         }
   \end{figure}

G280 is among three massive GCs in M31 for which colour-magnitude diagrams were published by F08. We  downloaded and reduced the same data (Programme ID 9719, P.I. T.\ Bridges), which were obtained with the High Resolution Channel (HRC) of the ACS.
The dataset includes four dithered exposures in each of the F814W and F606W filters with total integration times of 2860 s and 2020 s, respectively. 

We carried out point-spread function (PSF) fitting photometry on the drizzle combined images with the IRAF version of DAOPHOT~II \citep{Stetson1987}.  A first iteration of object detection on the F606W image was performed with the \texttt{daofind} task, followed by aperture photometry (\texttt{phot}), PSF construction (\texttt{psf}) and PSF-fitting photometry (\texttt{allstar}). Additional stars were then detected on the residual F606W image that was produced by \texttt{allstar} in the first iteration, and improved PSFs were determined from images from which all stars except the PSF stars themselves had been subtracted. A second iteration of PSF-fitting was then carried out with \texttt{allstar} using the combined star list as input. 
The photometry was calibrated by matching the PSF magnitudes to aperture photometry of the PSF stars in an $r=4$ pixels aperture, and applying the encircled energy corrections and VEGAMAG photometric zero-points from \citet{Sirianni2005}. We assumed a distance modulus of $(m-M)_0 = 24.47$ \citep{McConnachie2004} and a foreground extinction in the ACS bands of $A_\mathrm{F606W} = 0.153$~mag and $A_\mathrm{F814W} = 0.095$~mag \citep[][via the NASA/IPAC Extragalactic Database]{Schlafly2011}.
   
Figure~\ref{fig:cmd} shows the resulting colour-magnitude diagram for stars located at radii between 30 pixels ($0\farcs75$) and 200 pixels ($5\arcsec$) from the centre of the cluster. Together with the data for G280 we show photometry for the Milky Way globular cluster 47~Tuc (NGC~104) with red dots, taken from the ACS Survey of Galactic Globular Clusters \citep[ACSGCS;][]{Sarajedini2007}.  A foreground extinction of $A_\mathrm{F606W} = 0.091$~mag and $A_\mathrm{F814W} = 0.056$~mag \citep{Schlafly2011} and a distance of 4.7 kpc \citep{Woodley2012} were assumed for 47~Tuc. We also plot Dartmouth isochrones \citep{Dotter2007} for an age of 11 Gyr and metallicities of $\mathrm{[Fe/H]}=-1.4$, $-1.0$, $-0.7$, and $-0.4$ and a MIST isochrone \citep{Choi2016,Dotter2016} for an age of 10 Gyr and $\mathrm{[Fe/H]}=-0.5$. It should be noted that the MIST isochrone is for solar-scaled abundance patterns, whereas the Dartmouth isochrones are $\alpha$-enhanced.

Most of the RGB stars in G280 are somewhat redder than those in 47~Tuc and the dispersion in the colours on the RGB appears to be larger than the photometric errors, which are indicated by the horizontal error bars in Fig.~\ref{fig:cmd}. This is in agreement with the analysis by F08, who also found the colour dispersion to be larger than the photometric errors as determined from artificial star experiments. F08 did not quote a mean metallicity but their CMD shows their mean fiducial line for the RGB stars in G280 falling between those of 47~Tuc and NGC~5927. According to the 2010 version of the \citet{Harris1996} catalogue, these two clusters have $\mathrm{[Fe/H]}=-0.72$ and $\mathrm{[Fe/H]}=-0.49$, respectively, so the CMD for G280 suggests a metallicity between these two values, which is consistent with previous spectroscopic studies.

\section{Chemical abundance analysis}

   \begin{figure}
   \centering
   \includegraphics[width=\columnwidth]{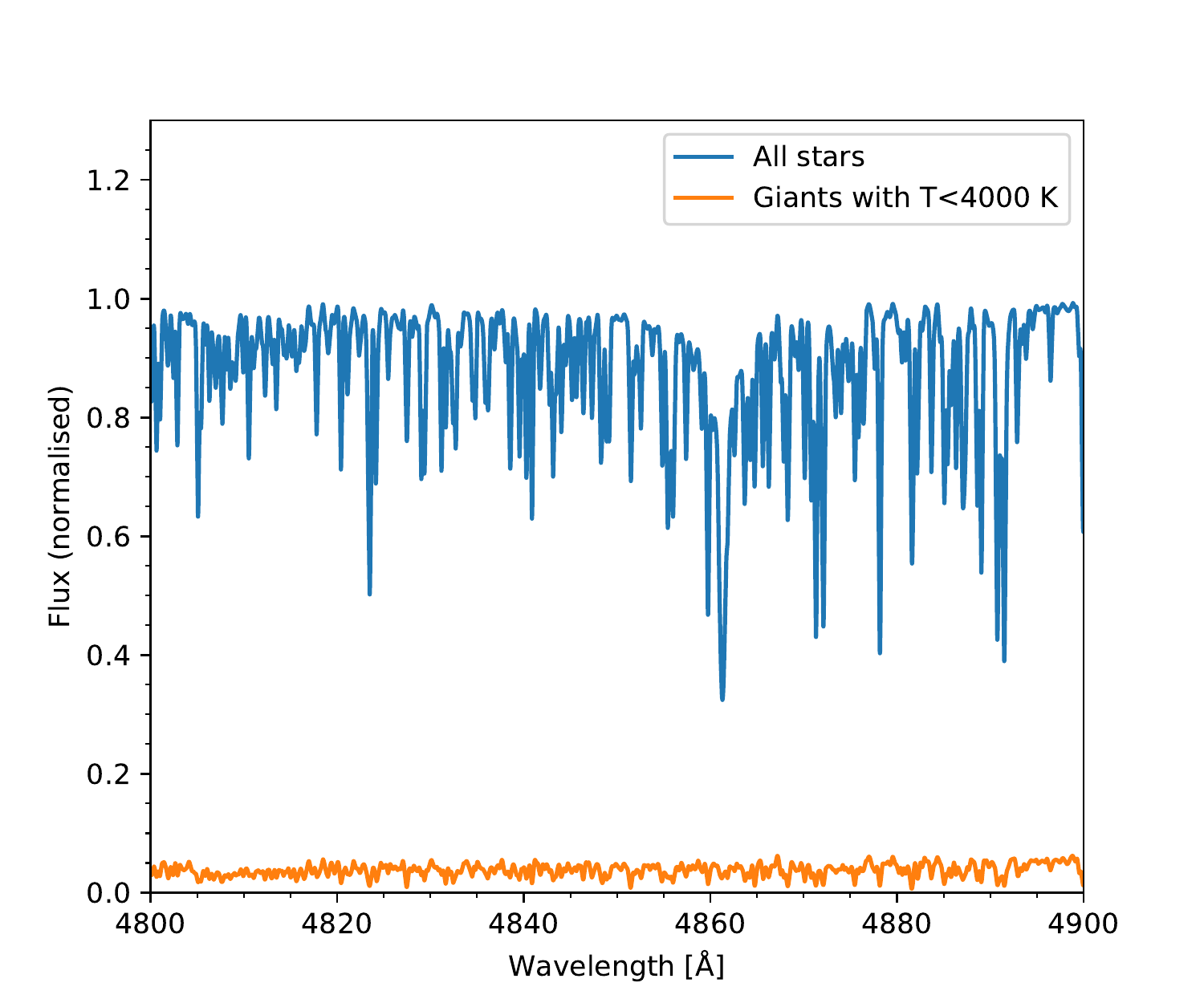}
   \includegraphics[width=\columnwidth]{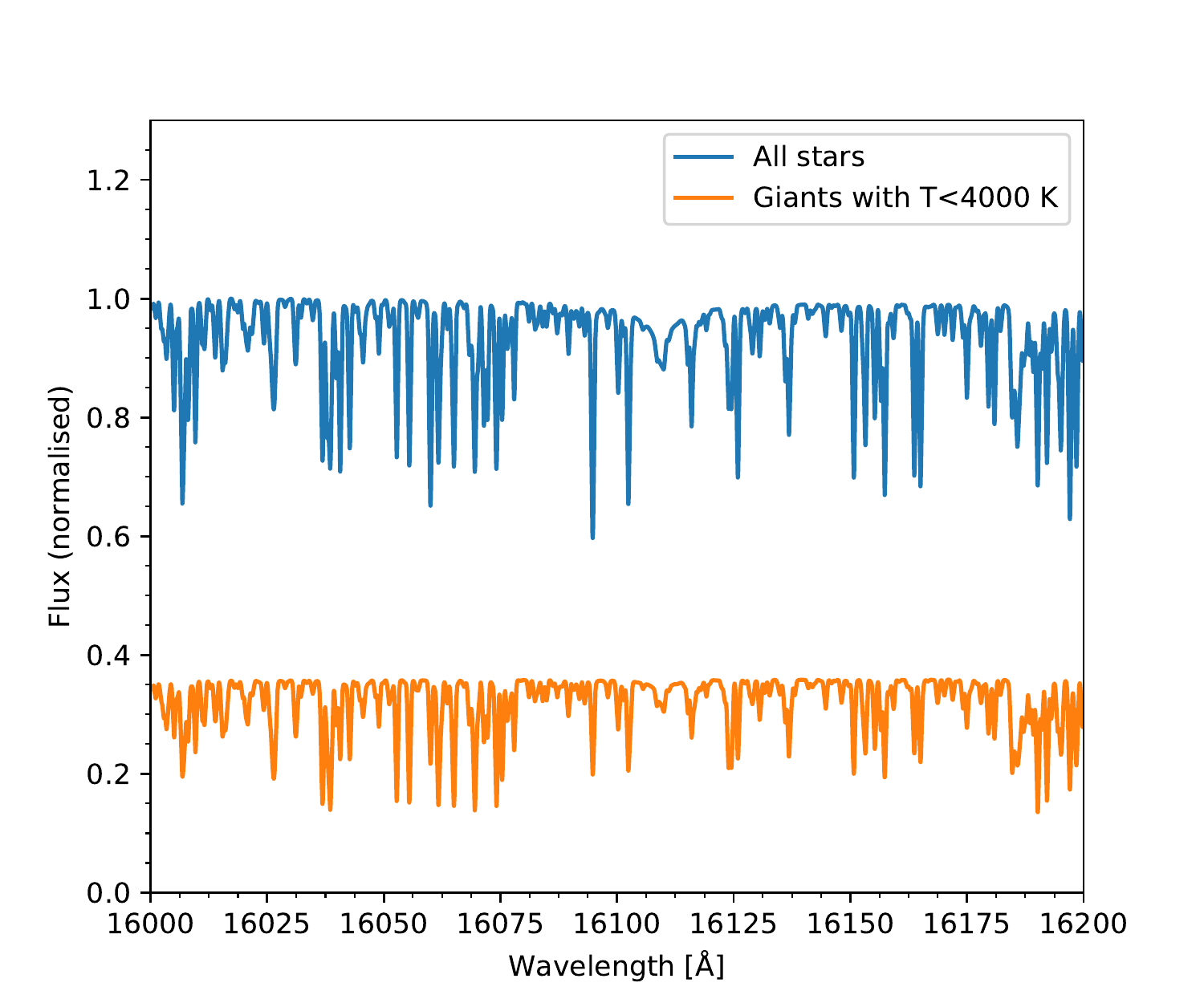}
      \caption{Contributions to the integrated light for all stars and for giants with $T_\mathrm{eff}<4000$ K.
      The top panel shows a small region of the optical range, whereas the bottom panel shows part of the $H$-band. The model spectra were calculated using Dartmouth isochrones with an age of 11 Gyr and a metallicity of  $\mathrm{[Fe/H]} = -0.70$.
         \label{fig:lfrac}
         }
   \end{figure}

   \begin{figure}
   \centering
   \includegraphics[width=\columnwidth]{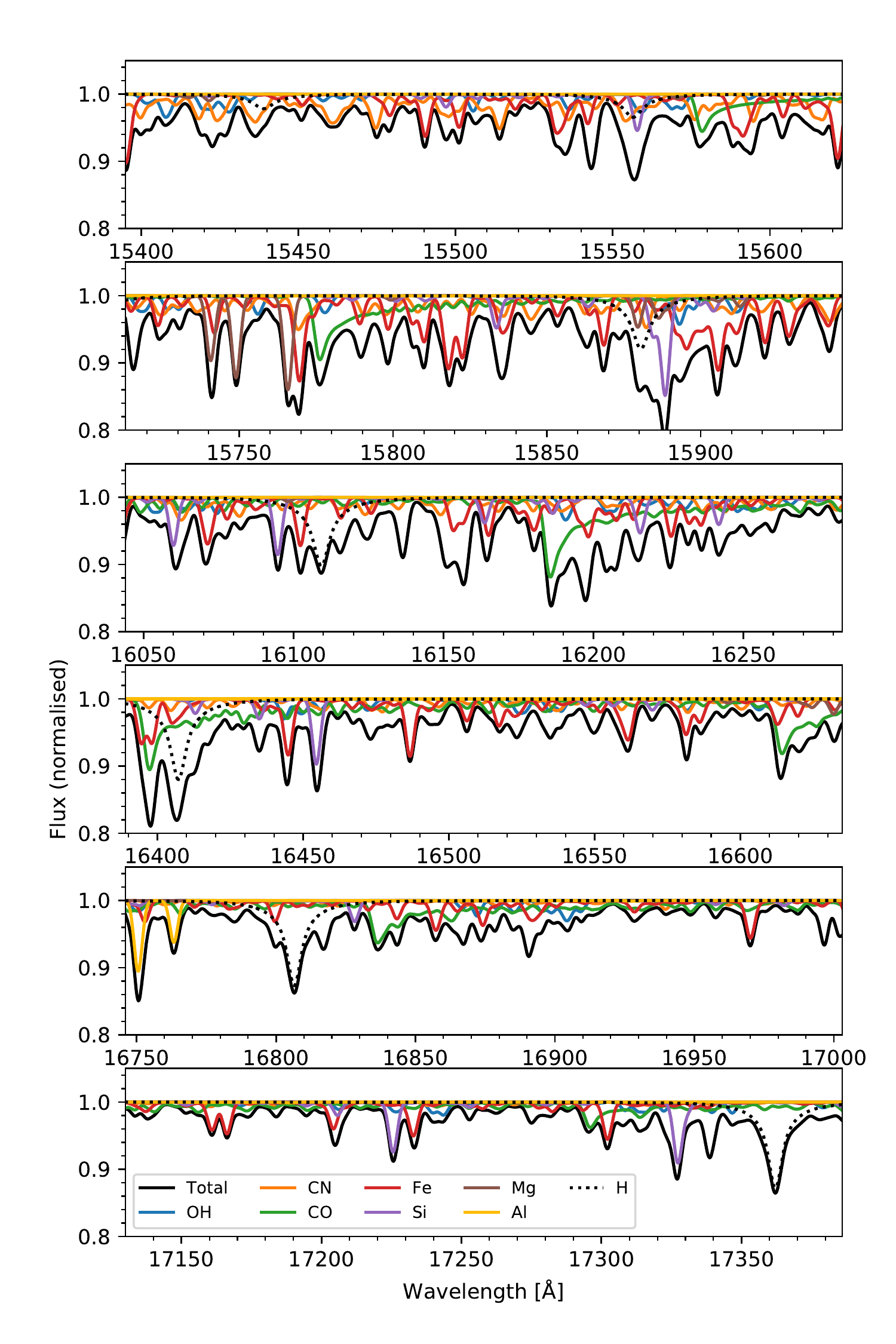}
      \caption{H-band model spectra for an Arcturus-like star for a number of the stronger features.
         \label{fig:synt}
         }
   \end{figure}

To measure chemical abundances from the spectra we proceeded as described in our previous papers \citep{Larsen2012a,Larsen2014,Larsen2017}. 
In the following we summarise the main aspects of our analysis procedure.

\subsection{General overview}

In short, abundances were derived by calculating simple stellar population (SSP) models, based on synthetic spectra, in which the abundances were iteratively adjusted until the best fit to the observed spectra (as determined by the $\chi^2$ of the residuals) was achieved.
To compute integrated-light model spectra, we first divided the HRD into about 90 bins (hereafter hrd-bins). Setting up the HRD is done independently of the actual fitting procedure, and may be based on empirical CMDs or on theoretical isochrones, or on a combination of both. For empirical CMDs, the temperatures and luminosities of the stars in each hrd-bin are typically determined from bolometric corrections and colour-$T_\mathrm{eff}$ transformations based on Kurucz model atmospheres \citep{Castelli2003}. Upon adopting a mass for each hrd-bin (e.g.\ based on the position of the star in the HRD), the surface gravity then follows from the standard relation,
\begin{equation}
\log g = \log g_\odot + \log \left[ \left(\frac{T_\mathrm{eff}}{T_\mathrm{eff,\odot}}\right)^4 \left(\frac{M}{M_\odot}\right) \left(\frac{L}{L_\odot}\right)^{-1} \right]
\end{equation}
for mass $M$, luminosity $L$, and effective temperature $T_\mathrm{eff}$.
In this paper, most of the modelling was based on theoretical isochrones, and stellar parameters could then be taken directly from the isochrone tables. Each hrd-bin was then assigned a weight $\mathscr{W}_i$, based on an assumed mass function $\phi(M) \propto \mathrm{d} N / \mathrm{d} M$, and given by
\begin{equation}
 \mathscr{W}_i = \phi(\langle M\rangle_i) \Delta M_i,
\end{equation}
where $\langle M \rangle_i$ is the average initial mass of stars in the $i$th bin and $\Delta M_i = \left| \langle M \rangle_{i+1} - \langle M \rangle_i\right|$.

Having defined the hrd-bins, model spectra were then computed for the corresponding stellar parameters of each hrd-bin, scaled by the weights $\mathscr{W}_i$, and co-added to produce a model spectrum for the cluster.
For most of the hrd-bins, we used the \texttt{ATLAS9} and \texttt{SYNTHE} codes \citep{Kurucz1970,Kurucz1979,Kurucz1981,Kurucz2005,Sbordone2004} to synthesise the model spectra, but for the coolest giants ($T_\mathrm{eff} < 4000$ K) we used \texttt{MARCS} atmospheres \citep{Gustafsson2008} and the \texttt{TurboSpectrum} code \citep{Alvarez1998,Plez2012} which allow for spherical symmetry.

For the microturbulence we used the same parameters as in our previous work \citep{Larsen2012a,Larsen2014,Larsen2017}:
for giants with $\log g < 1.0$ we assumed a microturbulent velocity of $\xi=2.0$ km s$^{-1}$ \citep[e.g.][]{Kraft1992} and for stars with $\log g > 4.5$ we assumed $\xi=0.5$ km s$^{-1}$ \citep{Takeda2002}. For the range $4.0 < \log g < 4.5$ we assumed $\xi=1$ km s$^{-1}$ \citep{Carretta2004}, and between $\log g = 4.0$ and $\log g = 1.0$ we interpolated linearly in $\log g$ between $\xi=1$ km s$^{-1}$ and $\xi=2$ km s$^{-1}$ \citep[see also][]{McWilliam2008}.
For stars on the horizontal branch we assumed $\xi=1.8$~km~s$^{-1}$ \citep{Pilachowski1996}.

The synthetic spectra were initially computed at very high resolution to sample the line profiles well ($\lambda/\Delta\lambda = 5\times10^5$, where $\Delta \lambda$ is the wavelength step per pixel). The actual resolution was dominated by the velocity broadening of the cluster, which was assumed to be well represented by a Gaussian distribution of line-of-sight velocities. An approximately Gaussian velocity distribution is expected on theoretical grounds since GCs are generally well-relaxed systems \citep{King1966} and is also largely supported by observations, although rotation may induce minor departures \citep{Heyl2012,Heyl2017}.

Because the HIRES and NIRSPEC spectra were not flux calibrated, it was necessary to apply a wavelength-dependent scaling of the flux level as part of the fitting procedure. For fits to individual spectral features, which were typically carried out in 10~\AA\ wide windows, we used a linear scaling as a function of wavelength (i.e.\ fitting only for the overall scaling and slope of the spectra). For fits extending over an entire echelle order we used a cubic spline, and for intermediate cases we used second- or third order polynomial fits.

Once the scaling was determined, the $\chi^2$ of the fit was evaluated, followed by adjustment of the input abundances, and computation of a new set of synthetic spectra. This procedure was repeated until the best fit was obtained. For fits involving only a single element, a golden section search was used to find the minimum $\chi^2$.  For simultaneous fits of multiple elements, the downhill simplex algorithm was used \citep{Nelder1965,Press1992}.

After determining the radial velocity shift, 
we first carried out an initial set of fits in which we allowed the broadening of the spectra and the overall scaling of the abundance scale to vary as free parameters. 
We then proceeded to solve for individual elements, using the same wavelength bins as in our previous studies. For Fe we fit each echelle order one by one, while for most other elements we used smaller wavelength bins covering individual features/multiplets. For the optical spectrum the elements were fitted individually, but in the $H$-band we also carried out fits in which the abundances of several elements were allowed to vary simultaneously.

Our procedure has been thoroughly described and tested in previous papers, in particular in \citet[][hereafter L17]{Larsen2017} where we compared integrated-light measurements of Galactic GCs with data for individual stars. \citet{Hernandez2017} applied the same methodology to an X-SHOOTER spectrum of the young cluster NGC~1705-1 and found good agreement between metallicity determinations from the $J$-band and the optical part of the spectrum. However, the present study is our first attempt to extend the analysis further to the infrared for an old GC. 

The CMD in Fig.~\ref{fig:cmd} is too crude to be used as a basis for modelling of the integrated-light spectra. Instead, we carried out several sets of fits, using theoretical Dartmouth and MIST isochrones and the empirical CMD of 47~Tuc. 
A discussion of the sensitivity of the results to the details of HRD modelling follows below (Sec~\ref{sec:diffhrd}).

\subsection{Line lists}

The line list is a critical part of any abundance analysis. Our previous work relied on the list by \citet[hereafter CH04]{Castelli2004} with the modifications described in L17, i.e.\ adding hyperfine splitting for Mn and Sc and updating the oscillator strengths for several Mg and Ba lines.
The tests described in L17 used this list, but only covered the optical region (4200~\AA -- 6200~\AA) of the spectra. The CH04 list is based on an older version of the line list available via the Kurucz web site. In the meantime, the Kurucz list has undergone many recent revisions, and in this paper we compare fits based on the CH04 list with the most recent version of the Kurucz atomic line list\footnote{http://kurucz.harvard.edu/linelists.html} available at the time the analysis was carried out (dated 8 Oct 2017).
To anticipate the later discussion, the results change relatively little for the optical part of the spectrum whereas more significant differences are seen in the $H$-band. It should be noted that whereas \texttt{TurboSpectrum} supports the use of damping constants from \citet{Barklem2000}, these are not included in the Kurucz list so for the \texttt{TurboSpectrum} calculations we fell back on the radiative and van der Waals damping constants from the Kurucz list. These damping constants are also used by \texttt{SYNTHE}. The damping constants for the \ion{Ba}{ii} lines are missing in the on-line version of the Kurucz list, so we added them from the CH04 list. Furthermore, we modified the isotopic mixture for Ba to match the $r$-process dominated composition according to \citet{McWilliam1998}, as discussed in L17. 
For the modelling of CN and CH we used the line lists from \citet{Masseron2014} and \citet{Brooke2014}.
In all cases, abundances are given relative to the solar composition according to \citet{Grevesse1998}.

\subsection{Remarks on modelling of the $H$-band spectrum}

While no single stellar type dominates the integrated $H$-band light for an old GC, the contribution from cool giants is clearly more important at infrared wavelengths than in the optical. This is illustrated quantitatively in Fig.~\ref{fig:lfrac} which shows model spectra for a population with an age of 11 Gyr and $\mathrm{[Fe/H]}=-0.70$, which is approximately representative of G280. We show spectra for the optical region near H$\beta$ and for a region near the centre of the $H$-band, near 1.6~$\mu$m.  Each panel shows the full normalised integrated-light spectrum and the contribution from giants with $T_\mathrm{eff}<4000$ K. In the optical part of the spectrum (upper panel), these cool giants only contribute $\sim5$\% of the flux, whereas they account for $\sim$34\% of the flux in the $H$-band (lower panel). While the temperature cut is somewhat arbitrary, it is clear that accurate modelling of the coolest giants becomes much more critical in the infrared.

In the optical, the abundance of Fe is usually well constrained by the large number of Fe lines. The $H$-band spectrum, instead, is dominated by strong molecular absorption features in the cool stars that contribute a large fraction of the flux at infrared wavelengths. To illustrate the contribution of various absorbers to the $H$-band spectrum, we computed synthetic spectra for an Arcturus-like star ($T_\mathrm{eff} = 4286$ K, $\log g = 1.66$, $\mathrm{[Fe/H]} = -0.5$) for various molecules and elements. In Fig.~\ref{fig:synt}, we show model spectra that include the OH, CN, and CO molecules and atomic transitions from Fe, Si, Mg, Al, and H for the wavelength intervals that correspond to the coverage of our NIRSPEC data. We also show synthetic spectra that include the full set of molecular and atomic transitions. The model spectra were convolved with a Gaussian kernel corresponding to the velocity dispersion of G280.
The $H$-band spectra include absorption bands from CO, OH and (mainly towards the bluer orders) CN and a number of Fe lines. Several strong absorption lines from Si, Mg, Al, and the hydrogen Brackett series are also present. Owing to the high velocity dispersion of G280, most features are strongly blended. 

Parts of the $H$-band are affected by telluric absorption features. Carrying out a full and adequate correction for these is notoriously difficult, and generally requires that standard stars are observed close in time and near the science target on the sky \citep[although progress is being made in theoretical modelling of the telluric absorption, e.g.][]{Smette2015,Gonneau2016}.  While a standard star was observed during our NIRSPEC run, it turned out to be insufficient for accurate telluric correction. We therefore only used the standard star to flag regions of the spectrum that are significantly affected by telluric lines, and then masked out these regions (by assigning zero weight to them) in the analysis.
To identify the regions affected by telluric absorption lines, we first median filtered the spectrum of the telluric standard, HR~1123 (a B0.5V star) with a 21 pixels wide median filter. This filtering eliminated the narrow telluric absorption features, while preserving broader features intrinsic to the standard star spectrum (mainly hydrogen Brackett lines). The median filtered spectrum was then subtracted from the original spectrum, and pixels with a deficit of more than 5\% of the original counts were flagged as affected by telluric absorption.

In addition to the telluric absorption features, the $H$-band spectra are affected by OH sky emission lines. Since these are variable on short time scales ($\sim$ minutes), they tend to leave residuals when the A-B exposure pairs are subtracted from each other. However, the affected pixels also have larger associated errors, so that they receive correspondingly lower weights in the fitting process.  

\section{Results}

   \begin{figure}
   \centering
   \includegraphics[width=\columnwidth]{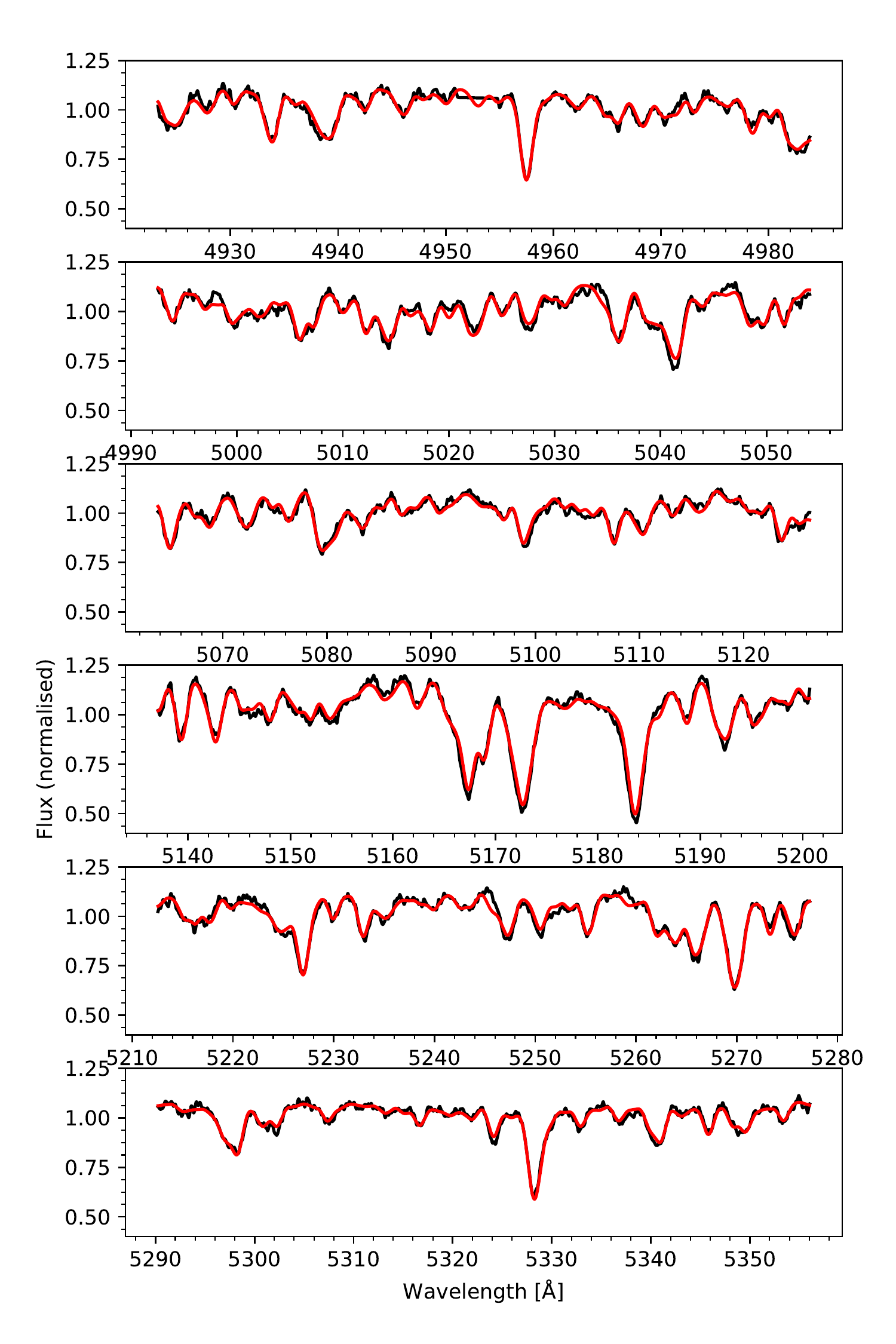}
      \caption{Optical spectrum of G280 and best-fitting models. Both the observed and model spectra are convolved with a Gaussian kernel with a dispersion of 2 pixels.
         \label{fig:fitopt}
         }
   \end{figure}

   \begin{figure}
   \centering
   \includegraphics[width=\columnwidth]{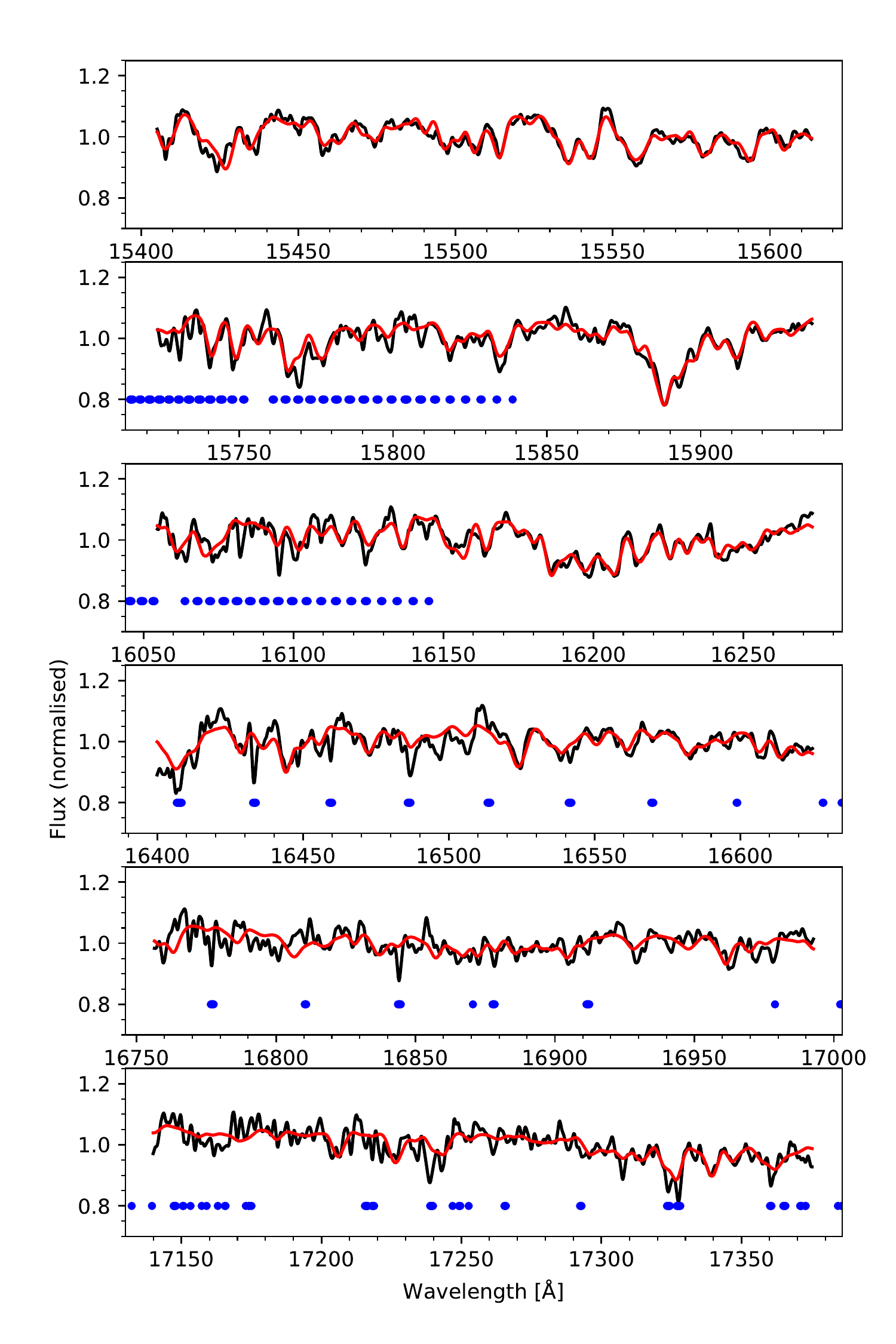}
      \caption{H-band spectrum of G280 and best-fitting models. Both the observed and model spectra are convolved with a Gaussian kernel with a dispersion of 2 pixels. The blue markers indicate regions of the spectra that are significantly affected by telluric absorption.
         \label{fig:fitir}
         }
   \end{figure}

   \begin{figure}
   \centering
   \includegraphics[width=\columnwidth]{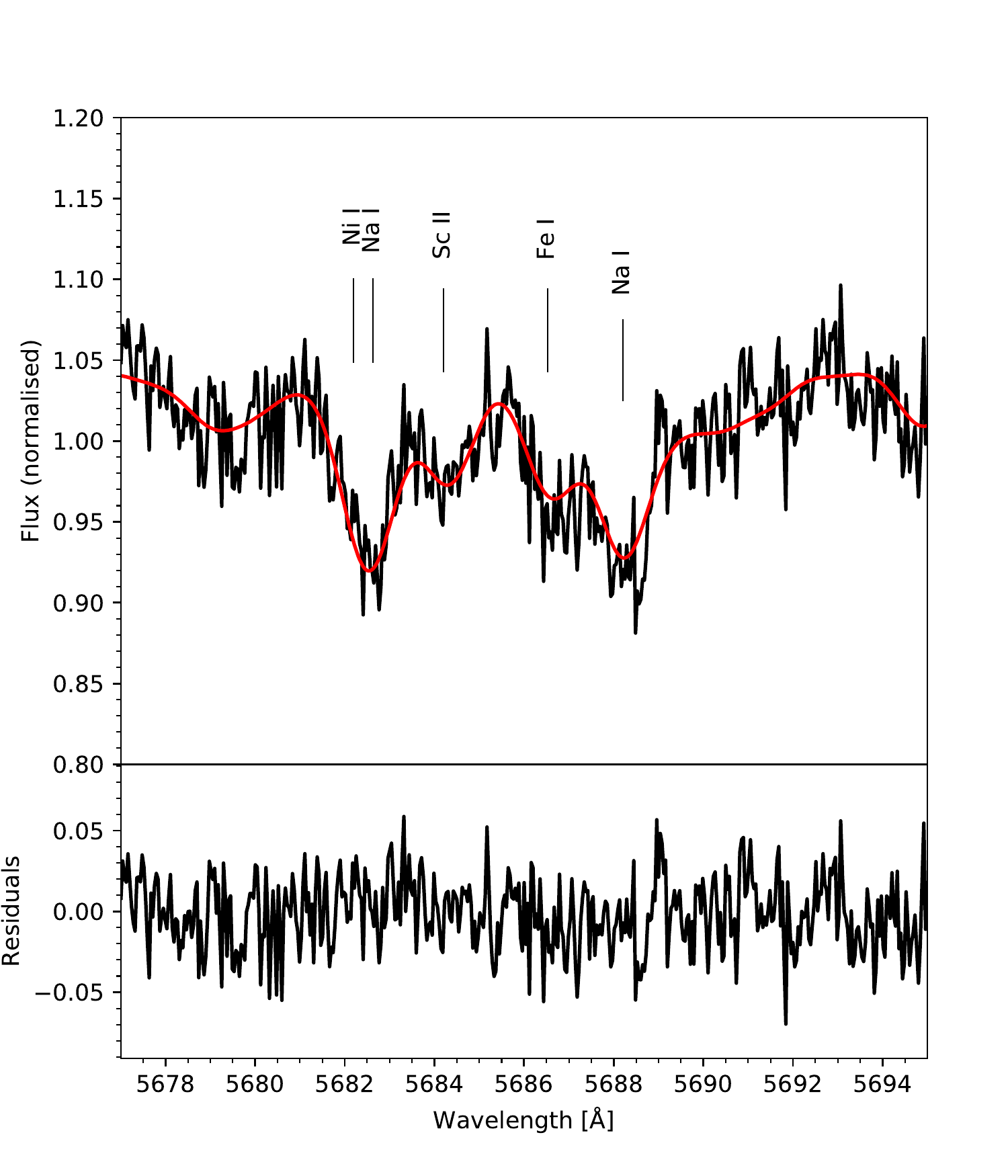}
      \caption{Fit to the \ion{Na}{i} doublet near 5683/5688~\AA. The bottom panel shows the (data - fit) residuals.
         \label{fig:speccmp_na}
         }
   \end{figure}

\begin{table}
\caption{Average abundances from the HIRES spectrum.}
\label{tab:abun-opt}
\centering
\begin{tabular}{l c c c}
\hline\hline
Element                     & Value & rms$_w$ & $N$ \\ \hline
$\mathrm{[Fe/H]}$ & $-0.678$ & 0.128 & 29 \\
$\mathrm{[Ti/Fe]}$ & $+0.369$ & 0.130 & 12 \\
$\mathrm{[Ca/Fe]}$ &  $+0.334$ & 0.256 & 8 \\
$\mathrm{[Mg/Fe]}$ &  $+0.569$ & 0.115 & 3 \\
$\mathrm{[Na/Fe]}$ & $+0.565$ & 0.051 & 1 \\
$\mathrm{[Mn/Fe]}$ &  $-0.080$ & 0.041 & 1 \\
$\mathrm{[Sc/Fe]}$ &  $+0.313$ & 0.074 & 6 \\
$\mathrm{[Cr/Fe]}$ &  $+0.013$ & 0.270 & 11 \\
$\mathrm{[Ba/Fe]}$ &  $-0.211$ & 0.111 & 3 \\
$\mathrm{[Ni/Fe]}$ &  $+0.197$ & 0.262 & 11 \\
\hline
\end{tabular}
\tablefoot{The fits were carried out using a Dartmouth isochrone with an age of 11 Gyr, a metallicity of $\mathrm{[Fe/H]}=-0.7$, an $\alpha$-enhancement of  $[\alpha/\mathrm{Fe}]=+0.4$,  the horizontal branch from 47~Tuc, and the Kurucz line list. See text for details.}
\end{table}

\begin{table}
\caption{Average abundances from the NIRSPEC spectrum from element-by-element fits.}
\label{tab:abun-ir1}
\centering
\begin{tabular}{l c c c}
\hline\hline
Element                     & Value & rms$_w$ & $N$ \\ \hline
$\mathrm{[Fe/H]}$ & $-0.598$ & 0.148 & 6 \\
$\mathrm{[C/Fe]}$ & $+0.070$ & 0.123 & 6 \\
$\mathrm{[O/Fe]}$ &  $+0.541$ & 0.174 & 2 \\
$\mathrm{[Mg/Fe]}$ &  $+0.163$ & 0.106 & 1 \\
$\mathrm{[Si/Fe]}$ &  $+0.355$ & 0.049 & 4 \\
\hline
\end{tabular}
\tablefoot{Details of the fits are the same as indicated in Table~\ref{tab:abun-opt}.}
\end{table}

\begin{table}
\caption{Average abundances from the NIRSPEC spectrum from simultaneous fitting of all elements. }
\label{tab:abun-ir2}
\centering
\begin{tabular}{l c c c}
\hline\hline
Element                     & Value & rms & $N$ \\ \hline
$\mathrm{[Fe/H]}$ & $-0.647$ & 0.171 & 6 \\
$\mathrm{[C/Fe]}$ & $-0.112$ & 0.262 & 6 \\
$\mathrm{[O/Fe]}$ &  $+0.443$ & 0.279 & 6 \\
$\mathrm{[Si/Fe]}$ &  $+0.334$ & 0.294 & 4 \\
\hline
\end{tabular}
\tablefoot{Details of the fits are the same as indicated in Table~\ref{tab:abun-opt}.}
\end{table}

From the HIRES spectrum we measured a heliocentric radial velocity of  $-161.8$ km~s$^{-1}$, which agrees with the values found by \citet{Strader2011} and \citet{Sakari2016} within about one km~s$^{-1}$. The total broadening derived from the HIRES spectrum corresponds to $\sigma_\mathrm{opt} = 26.9\pm0.3$ km~s$^{-1}$, where the error is the uncertainty on the mean determined from the standard deviation of the measurements from the 29 individual orders. Corrected for the instrumental resolution, $R=40\, 000$ or 3.2 km~s$^{-1}$, this gives an intrinsic line-of-sight velocity dispersion for the cluster spectrum of 26.7 km~s$^{-1}$, which is close to the velocity dispersion of $\sigma = 25.94$ km~s$^{-1}$ from \citet{Djorgovski1997}. This may be compared with the central velocity dispersion of  $\sigma_0  = 32.0\pm1.7$ km~s$^{-1}$ and the global dispersion of $\sigma_\infty = 24.9\pm1.3$ km~s$^{-1}$ obtained by \citet{Strader2011}. From the NIRSPEC data we got an average broadening of $\sigma_\mathrm{IR} = 29.7 \pm 1.8$ km~s$^{-1}$, which gives an intrinsic velocity dispersion of 29.2 km~s$^{-1}$ when corrected for the instrumental resolution of $R=25\, 000$ (5.1 km~s$^{-1}$). The larger uncertainty here is due to the smaller number of orders and a greater order-to-order dispersion. The difference between the velocity dispersions derived from the HIRES and NIRSPEC data may be partly real, since the NIRSPEC slit was only half as wide as the HIRES slit and thus preferentially sampled the central regions of the cluster where the velocity dispersion is expected to be higher. However, since the velocity dispersion derived from the NIRSPEC data is consistent with that derived from the HIRES data within about 1.3~$\sigma$, but has a larger uncertainty, we adopted
the same broadening for the NIRSPEC data as for the HIRES data, adding the difference between the instrumental resolutions in quadrature. In Sect.~\ref{sec:dsig} we comment on the sensitivity of the results to the adopted broadening.

\subsection{Reference analysis: Kurucz line list, Dartmouth isochrone}

Our first set of fits, which are used as a reference in the rest of this paper, used Dartmouth isochrones for an age of 11 Gyr and $\alpha$-enhanced composition ($[\alpha/\mathrm{Fe}]=+0.4$). The isochrones were combined with the empirical data for the horizontal branch of 47~Tuc from the ACSGCS photometry. The appropriate weighting of the ACSGCS data was determined by scaling the isochrone weights such that the number of RGB stars predicted in the magnitude range $+1 < M_V < +2$ matched that in the ACSGCS data. Physical parameters for the HB stars were obtained using  colour-$T_\mathrm{eff}$ transformations and bolometric corrections based on  Kurucz models \citep{Castelli2003}.
As in our previous work, we included stars down to an absolute magnitude limit of $M_V=+9$ and the distribution of stellar masses was assumed to follow the \citet{Salpeter1955} form, $\mathrm{d}N/\mathrm{d} M \propto M^{-2.35}$. The metallicity of the isochrones was selected to self-consistently match that obtained from the fits (within 0.1 dex), in this case $\mathrm{[Fe/H]}=-0.7$. For these fits, the model spectra were synthesised using the Kurucz line list. 

Figure~\ref{fig:fitopt} shows a few of the orders for the HIRES spectrum along with the best-fitting models, and Fig.~\ref{fig:fitir} shows the corresponding comparison for the NIRSPEC data. For the latter, we also indicate the regions that are (strongly) affected by telluric lines and therefore were excluded from the fits. It is evident that telluric absorption features, which are recognizably sharper than the velocity-broadened features originating from G280 itself, are present at many of the indicated wavelengths. An advantage of the relatively high spectral resolution of an instrument such as NIRSPEC is thus that these regions can be masked out while still leaving pixels in between the telluric lines available for the spectral fitting.

In Table~\ref{tab:ab-opt} we list the abundances obtained for each wavelength bin for the HIRES data, and Table~\ref{tab:ab-ir} gives the same information for the NIRSPEC data. Table~\ref{tab:ab-irn} gives the measurements obtained by letting the abundances of multiple elements vary simultaneously for the NIRSPEC data. The corresponding average abundances are given in Tables~\ref{tab:abun-opt}--\ref{tab:abun-ir2}. For each element we also give the weighted rms dispersion of the individual measurements (rms$_w$) and the number of wavelength bins, $N$, where rms$_w$ was calculated as
\begin{equation}
 \mathrm{rms}_w = \left(\frac{\sum w_i \left(\mathrm{[X/Fe]}_i - \langle \mathrm{[X/Fe]} \rangle\right)^2}{\sum w_i}\right)^{1/2}
\end{equation}
for weights
\begin{equation}
w_i = \left[\sigma_i^2 + (0.01 \, \mathrm{dex})^2\right]^{-1}.
\end{equation}
We note that, as in L17, a ``floor'' of 0.01 dex was added to the formal uncertainties on the fitted values.
Taking rms$_w$ as an indication of the uncertainties on the individual measurements, we can then estimate the errors on the mean values as $\sigma = \mathrm{rms}_w / \sqrt{N-1}$. 

A limitation of the multi-element fits is that they do not provide error estimates on the fitted abundances.  The average abundances in Table~\ref{tab:abun-ir2} are therefore unweighted, and we give the unweighted rms instead of rms$_w$. For this reason, we will base the discussion mostly on the element-by-element fits.

The average iron abundance obtained from the optical data, $\mathrm{[Fe/H]}=-0.68$ with rms$_w=0.13$ dex, agrees very well with previous spectroscopic metallicity determinations \citep{Caldwell2011,Colucci2014,Sakari2016}. We also find that the $\alpha$-elements are enhanced compared to solar scaled abundance patterns, with a formal mean of $\langle [(\mathrm{Mg, Ca, Ti})/\mathrm{Fe}] \rangle = +0.40\pm0.03$. In spite of the relatively large velocity dispersion of G280, several weaker features, such as the \ion{Na}{i} doublet at 5683/5688~\AA , are quite readily detectable in our HIRES spectrum (Fig.~\ref{fig:speccmp_na}). The $\mathrm{[Na/Fe]}$ ratio is clearly super solar and higher than the $\mathrm{[Na/Fe]}$ ratios typically observed in field stars in the Milky Way at these metallicities, but it is very similar to the mean $\mathrm{[Na/Fe]}$ measured for metal-rich Galactic GCs (47~Tuc, NGC~6388) by \citet{Carretta2009}. The data thus suggest that G280, like most of its Galactic counterparts, hosts a significant population of substantially Na-enriched stars. Our Mg abundance, $\mathrm{[Mg/Fe]}=+0.57\pm0.08$ agrees with that measured by \citet{Colucci2014} ($\mathrm{[Mg/Fe]}=+0.45\pm0.1$) within the uncertainties. 

   \begin{figure}
   \centering
   \includegraphics[width=\columnwidth]{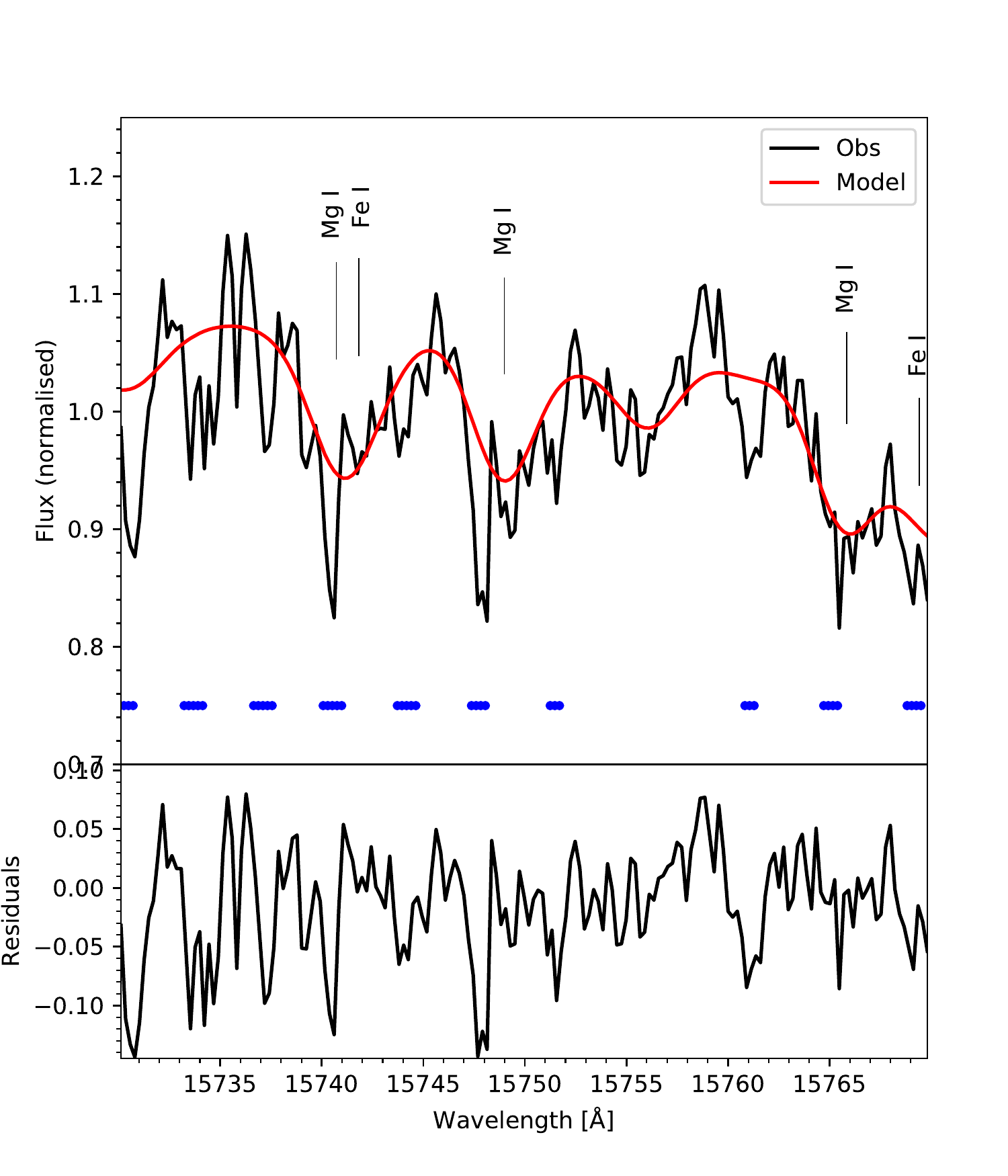}
      \caption{Fit to the \ion{Mg}{i} triplet near 15750~\AA. The blue markers indicate regions affected by telluric absorption and the bottom panel shows the (data - fit) residuals.
         \label{fig:speccmp_mg}
         }
   \end{figure}

   \begin{figure}
   \centering
   \includegraphics[width=\columnwidth]{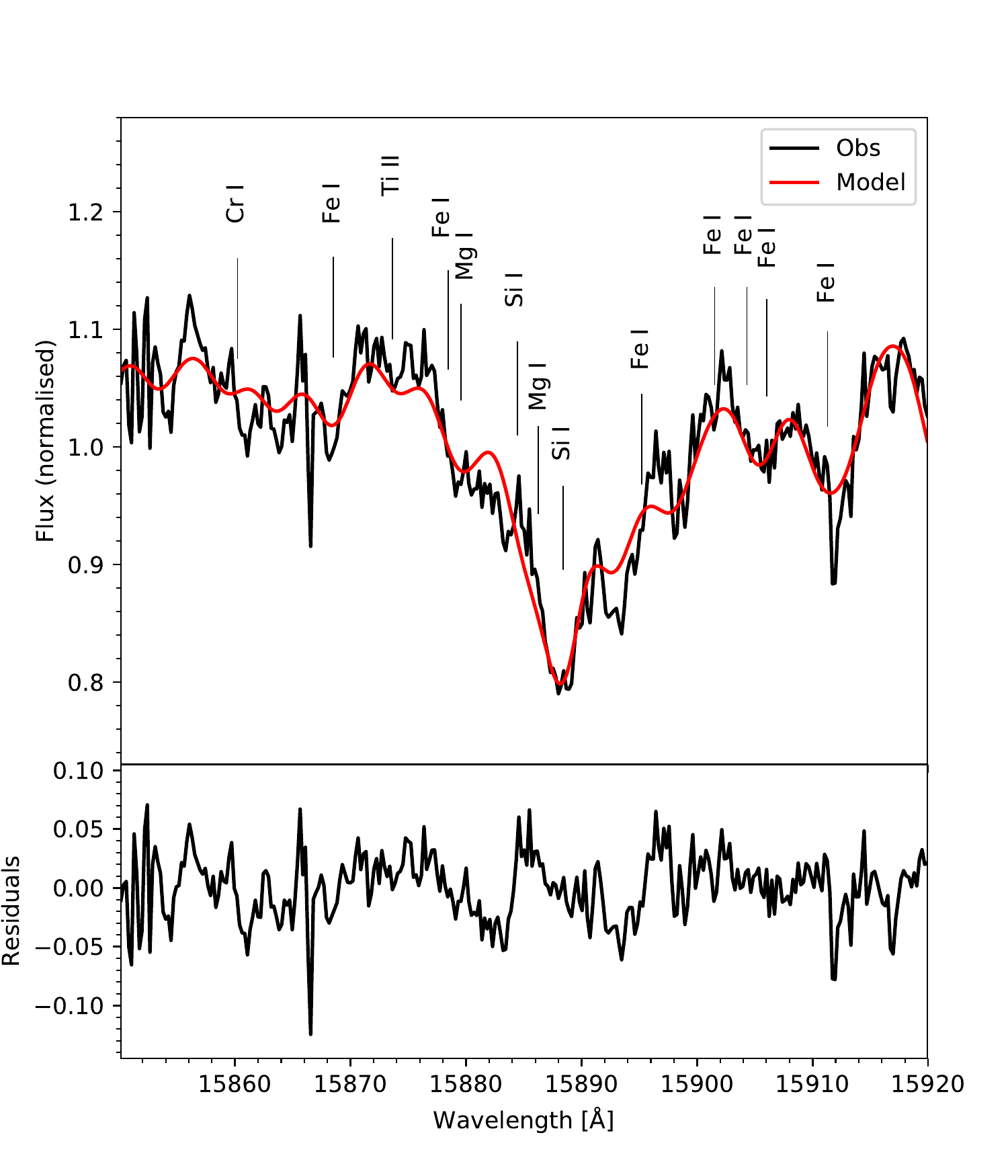}
      \caption{Fit to the \ion{Si}{i} line at 15888~\AA. The bottom panel shows the (data - fit) residuals.
         \label{fig:speccmp_si}
         }
   \end{figure}

For these fits, the Fe abundance derived from the NIRSPEC data, $\mathrm{[Fe/H]}=-0.60$ with rms$_w = 0.15$ dex, agrees well with that obtained from the HIRES data, with a  slightly higher order-to-order rms dispersion for the NIRSPEC data. Given the uncertainties, the 0.08 dex difference between the average HIRES and NIRSPEC iron abundance measurements from the element-by-element fits is insignificant.
The iron abundance from the multi-element NIRSPEC fits is in even better agreement with that obtained from the HIRES data, but this is mainly due to the difference between weighted (for the element-by-element fits) and unweighted (for the multi-element fits) averages. Indeed, a straight average of the element-by-element NIRSPEC fits gives $\mathrm{[Fe/H]}=-0.671$, very close to the value obtained from the multi-element fits.

Fits to individual $H$-band features are shown in Fig.~\ref{fig:speccmp_mg} (\ion{Mg}{i}) and Fig.~\ref{fig:speccmp_si} (\ion{Si}{i}).
The NIRSPEC data  yield a slightly lower average $\alpha$-element enhancement than that found from the HIRES data, albeit for a different set of $\alpha$-elements, with $\langle [(\mathrm{O, Mg, Si})/\mathrm{Fe}] \rangle = +0.35\pm0.03$. Our $\alpha$-element abundances measured from the NIRSPEC spectrum also agree well with those measured by \citet{Sakari2014}, who found  $\mathrm{[O/Fe]} = +0.39\pm 0.09$, $\mathrm{[Mg/Fe]} = +0.24\pm0.14$, and $\mathrm{[Si/Fe]} = +0.32\pm0.06$. 
While our Mg abundance measured from the $H$-band data is consistent with that of \citet{Sakari2014}, it is about 0.4 dex lower than that measured from our HIRES data, a difference that exceeds the formal errors of about 0.1 dex. The origin of this difference is not clear, but we note that the $H$-band measurement comes from a single window around the \ion{Mg}{i} triplet near 15750~\AA, which is blended with relatively strong Fe lines and CN bands (Fig.~\ref{fig:synt}) and is also affected by telluric absorption (Fig.~\ref{fig:speccmp_mg}). 

From the element-by-element fits we find a somewhat higher $\mathrm{[C/Fe]}$ ratio than \citet{Sakari2014} (who found $\mathrm{[C/Fe]} = -0.21 \pm 0.05$). The multi-element fits in Table~\ref{tab:abun-ir2} give a C abundance closer to (but still slightly higher than) that measured by \citet{Sakari2014}. 
Again, the difference between the C abundances in Table~\ref{tab:abun-ir1} and Table~\ref{tab:abun-ir2} is mainly due to the fact that the average values listed in Table~\ref{tab:abun-ir2} are unweighted.
The average C abundance in Table~\ref{tab:abun-ir1} is strongly dominated by the first echelle order, and a straight average of the element-by-element fits instead gives $\mathrm{[C/Fe]} = -0.08$,  in good agreement with the multi-element fit, but still about 0.1 dex higher than the measurement by \citet{Sakari2014}. In any case, we note that the carbon abundances measured from integrated-light spectra are affected by mixing of CNO-processed material along the RGB, which dilutes the C abundances for stars above the RGB bump \citep{Gratton2000,Martell2008}.

\subsection{Modified input assumptions}
\label{sec:modasump}

\begin{table}
\caption{Changes in abundances for different input assumptions}
\label{tab:abvar}
\centering
\begin{tabular}{l c c c c c}
\hline\hline
Delta   & CH04   & Kroupa & 47 Tuc & MIST & $\sigma\times1.1$ \\ \hline        
\hline
\multicolumn{2}{l}{HIRES}\\
\mbox{[Fe/H]} & $-0.020$ & $-0.008$ & $+0.025$ & $-0.087$ & $+0.065$ \\
\mbox{[Ti/Fe]} & $-0.058$ & $+0.008$ & $-0.002$ & $-0.013$ & $+0.038$ \\
\mbox{[Ca/Fe]} & $-0.075$ & $-0.035$ & $+0.014$ & $-0.011$ & $-0.020$ \\
\mbox{[Mg/Fe]} & $-0.083$ & $+0.004$ & $+0.028$ & $+0.094$ & $+0.001$ \\
\mbox{[Mn/Fe]} & $-0.062$ & $-0.003$ & $+0.017$ & $-0.042$ & $+0.065$ \\
\mbox{[Sc/Fe]} & $-0.082$ & $+0.022$ & $-0.064$ & $-0.055$ & $+0.047$ \\
\mbox{[Cr/Fe]} & $+0.016$ & $-0.017$ & $+0.017$ & $-0.017$ & $-0.017$ \\
\mbox{[Na/Fe]} & $+0.001$ & $-0.038$ & $-0.002$ & $+0.032$ & $+0.048$ \\
\mbox{[Ba/Fe]} & $+0.133$ & $+0.013$ & $-0.022$ & $-0.071$ & $+0.036$ \\
\mbox{[Ni/Fe]} & $+0.115$ & $+0.012$ & $-0.013$ & $+0.008$ & $+0.027$ \\
\multicolumn{2}{l}{NIRSPEC} \\
\mbox{[Fe/H]} & $+0.241$ & $+0.033$ & $+0.013$ & $-0.081$ & $+0.056$ \\
\mbox{[C/Fe]} & $-0.320$ & $+0.000$ & $-0.082$ & $-0.167$ & $-0.076$ \\
\mbox{[O/Fe]} & $-0.027$ & $-0.051$ & $+0.099$ & $-0.292$ & $-0.001$ \\
\mbox{[Mg/Fe]} & \ldots & $-0.003$ & $+0.006$ & $+0.067$ & $+0.051$ \\
\mbox{[Si/Fe]} & \ldots & $+0.022$ & $-0.025$ & $+0.020$ & $-0.017$ \\
\hline
\end{tabular}
\end{table}

To examine how much the modelling assumptions affect the results, we repeated the analysis for a variety of modified assumptions, as summarised in Table~\ref{tab:abvar}. The first column, labelled CH04, shows how the abundances change if we use the modified \citet{Castelli2004} line list instead of the Kurucz list. 
In the second column (``Kroupa''), we indicate how the abundances change when a Kroupa IMF extending to a lower mass limit of $0.2 \, M_\odot$ ($M_V\approx+12$) is adopted. Next (``47 Tuc''), we list the abundance changes when using the empirical CMD of 47~Tuc for the modelling. The column labelled ``MIST''  gives the changes for the MIST isochrone shown in Fig.~\ref{fig:cmd}, which has an age of 10 Gyr and  $\mathrm{[Fe/H]}=-0.5$ and includes stellar evolutionary phases up to the tip of the AGB. Finally, the last column (``$\sigma\times1.1$'') shows how the abundances are affected by an increase of 10\% in the dispersion of the Gaussian used to smooth the model spectra. Apart from the first column, all fits used the Kurucz line list.

\subsubsection{Different line list}

   \begin{figure}
   \centering
   \includegraphics[width=\columnwidth]{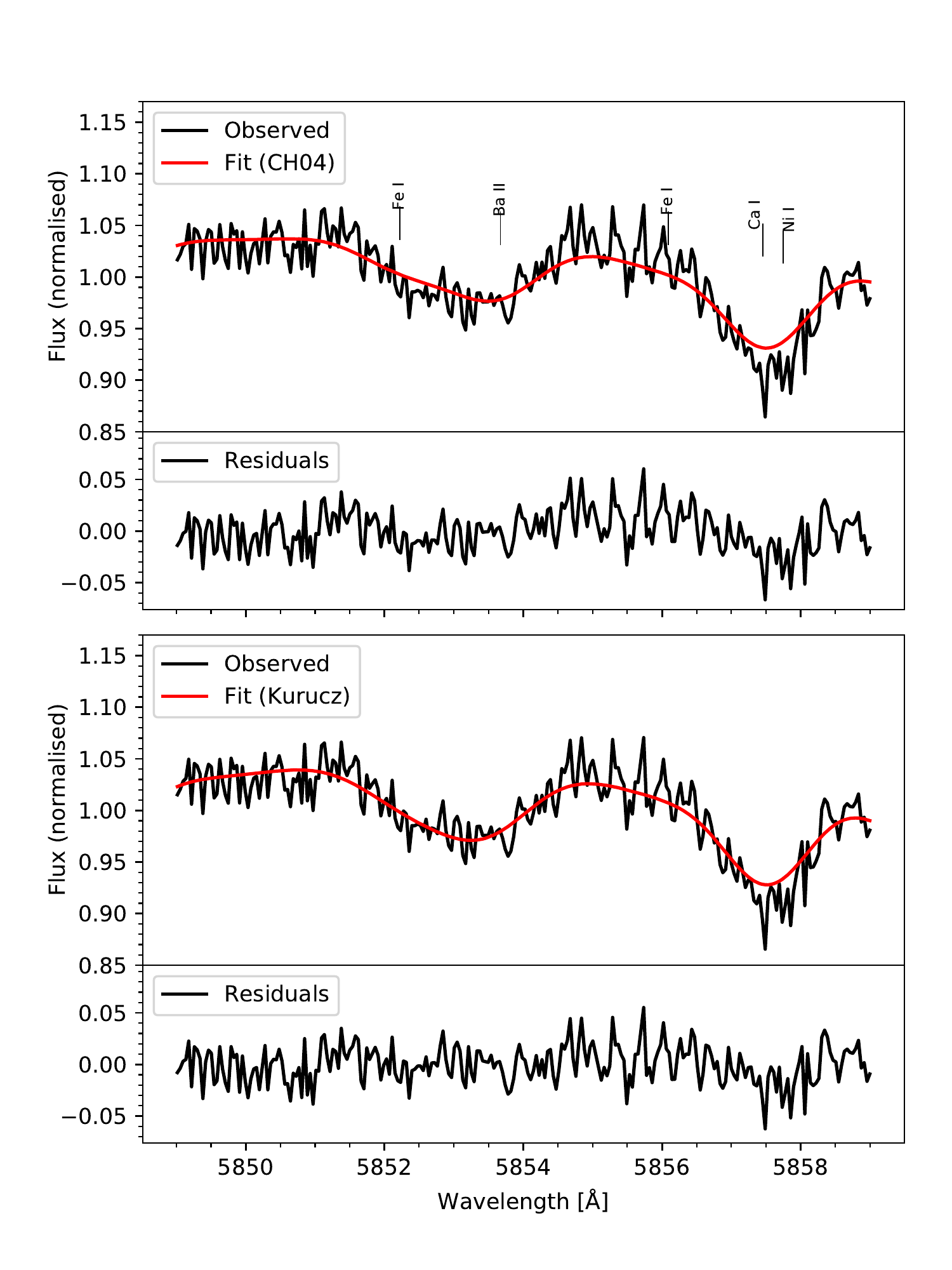}
      \caption{Fits to the \ion{Ba}{ii} line at 5854~\AA, using our modified version of the CH04 line list (top) and the Kurucz line list (bottom).
         \label{fig:bacmp}
         }
   \end{figure}

Our tests in L17 showed that the \citet{Castelli2004} line list with our modifications generally appeared adequate for analysis of optical spectra (4200~\AA -- 6200~\AA). Table~\ref{tab:abvar} confirms that most abundance ratios derived from the optical spectrum change relatively little ($<0.1$ dex) when the CH04 list is used instead of the Kurucz list, although changes greater than 0.1 dex are seen for Ba and Ni. For Ba, about 0.05 dex of the difference can be attributed to changes in the $\log gf$ values. The remainder is caused by other differences in the line lists that affect the fitting of the continuum level, especially affecting the 5854~\AA\ line. 
Visual inspection shows that the better fits are obtained with the Kurucz list, which is supported by the smaller rms line-to-line dispersion (rms$_w=0.111$ dex for the fits based on the Kurucz list versus 0.182 dex for the CH04 list). Figure~\ref{fig:bacmp} shows the fits to the \ion{Ba}{ii} line at 5854~\AA, which is blended with an \ion{Fe}{i} line at 5852.2~\AA\ at the velocity dispersion of G280. Although the differences are not dramatic, the Kurucz line list gives a better fit, which is confirmed by the better reduced $\chi^2$ value for the fit ($\chi^2_\nu=1.48$ for the Kurucz list versus $\chi^2_\nu=1.72$ for the CH04 list).
For Ni, the differences are caused by a combination of updated $\log gf$ values and inclusion of isotopic splitting for some lines, again combined with other differences in the line lists that affect the continuum scaling. 

More significant changes are seen in the infrared than in the optical. With the CH04 list we get an Fe abundance of $\mathrm{[Fe/H]} = -0.36$ from the NIRSPEC data with a high dispersion of rms$_w = 0.29$ dex, and the corresponding difference between the optical and near-infrared Fe abundance measurements is 0.34 dex.
The CH04 list does not include damping constants for the $H$-band Mg and Si lines, 
which prevents us from obtaining meaningful measurements of the abundances of these elements. Our inability to properly synthesise these transitions may also affect other abundances obtained from the $H$-band data with the CH04 list. 

\subsubsection{Kroupa IMF}

Table~\ref{tab:abvar} shows that the results change very little when the magnitude limit is increased from $M_V=+9$ to $M_V=+12$. A similar conclusion was reached by \citet{Larsen2012a} for the metal-poor clusters in the Fornax dwarf galaxy. This is quite reassuring, because the exact behaviour of the low-mass end of the stellar mass distribution in GCs may be affected by dynamical evolution, making it uncertain \citep[e.g.][]{Spitzer1987}. As a practical matter, including the faintest part of the main sequence would also pose additional challenges for the spectral synthesis owing to the low effective temperatures of these stars.

\subsubsection{Different HRDs}
\label{sec:diffhrd}

As noted in L17, the Dartmouth isochrones have difficulties matching the brightest part of the RGB of metal-rich clusters such as 47~Tuc and NGC~6388. From Fig.~\ref{fig:cmd} this is also true for G280. While an isochrone with $\mathrm{[Fe/H]}=-0.7$ can roughly reproduce the range in colours seen for the RGB stars, the tip of the RGB is too faint in the models. At lower metallicities, the tip of the RGB becomes brighter, but the isochrones do not extend as far to the red.

A more recent set of isochrones is MIST \citep{Choi2016,Dotter2016}, which includes stellar evolutionary phases beyond the tip-RGB, although the MIST isochrones are currently available for solar scaled composition only. As shown in Fig.~\ref{fig:cmd}, a MIST isochrone with $\mathrm{[Fe/H]}=-0.5$ provides a reasonable fit to the observed RGB/AGB of G280. It is important to note that the metallicity scale of the MIST isochrones is given relative to the solar composition from \citet{Asplund2009}, unlike the Dartmouth isochrones which assume the solar abundances in \citet{Grevesse1998}. Hence, a MIST isochrone with $\mathrm{[Fe/H]}=-0.5$ and $[\alpha/\mathrm{Fe}]=0$  corresponds to a metal fraction of $Z=4.64\times10^{-3}$, whereas a Dartmouth isochrone with $\mathrm{[Fe/H]}=-0.7$ and $[\alpha/\mathrm{Fe}] = +0.4$ has $Z = 6.51\times10^{-3}$. The total metal content of the $\mathrm{[Fe/H]}=-0.5$ solar scaled MIST isochrone is thus about 0.15 dex \emph{lower} than that of the $\mathrm{[Fe/H]}=-0.7$ $\alpha$-enhanced Dartmouth isochrone.

As noted in Section~\ref{sec:phot}, the metallicity of 47~Tuc may be slightly lower than that of G280, which is supported by the slightly bluer overall locus of the RGB of 47~Tuc (Fig.~\ref{fig:cmd}). Nevertheless, carrying out the abundance analysis for G280 using the CMD for 47~Tuc as input for the spectral modelling can still give some useful insight into the sensitivity of our results to the input assumptions. As can be seen from the figure, the RGB of the empirical CMD for 47~Tuc extends nearly as far to the red as that of G280. 

We see from Table~\ref{tab:abvar} that the abundance measurements are fairly robust with respect to changes in the assumptions about the HRD.  For both the 47~Tuc CMD and the MIST isochrone, the difference in $\mathrm{[Fe/H]}$ between the HIRES and NIRSPEC measurements remains very similar to that obtained from our reference fits: for the 47~Tuc CMD, we get $\mathrm{[Fe/H]}=-0.65$ (HIRES) and $\mathrm{[Fe/H]}=-0.59$ (NIRSPEC), and for the MIST isochrone we get $\mathrm{[Fe/H]}=-0.77$ (HIRES) and $\mathrm{[Fe/H]}=-0.68$ (NIRSPEC).
The largest changes occur for the $\mathrm{[O/Fe]}$ ratio derived from the NIRSPEC data, which changes by $+$0.10 dex when using the 47~Tuc CMD and by $-0.29$ dex for the MIST isochrone. This is probably a direct consequence of the fact that the near-infrared OH lines are very sensitive to variations in the stellar physical parameters, such as $T_\mathrm{eff}$ \citep{Smith2013}. Given that the OH lines are also relatively weak and blended with other features, it is perhaps not surprising that the oxygen abundances are more sensitive to changes in the model assumptions.

\subsubsection{Modifying the broadening}
\label{sec:dsig}

As noted earlier, we derived somewhat different velocity dispersions from the HIRES and NIRSPEC spectra, with the velocity dispersion derived from the NIRSPEC spectra formally being about 2.5~km~s$^{-1}$ higher.
While the significance of the difference is small, we carried out a set of fits in which the amount of broadening was increased by 10\%. This may be considered a generous estimate of the uncertainty, especially for the HIRES spectrum. Nevertheless, Table~\ref{tab:abvar} shows that the effect on the abundance determinations is modest. If we had adopted the formal average broadening derived from the NIRSPEC data, the $H$-band metallicity would have increased by about 0.05 dex, and the average $H$-band $[\alpha/\mathrm{Fe}]$ ratio would have decreased by $\sim0.01$ dex.

\section{Discussion}

   \begin{figure}
   \centering
   \includegraphics[width=\columnwidth]{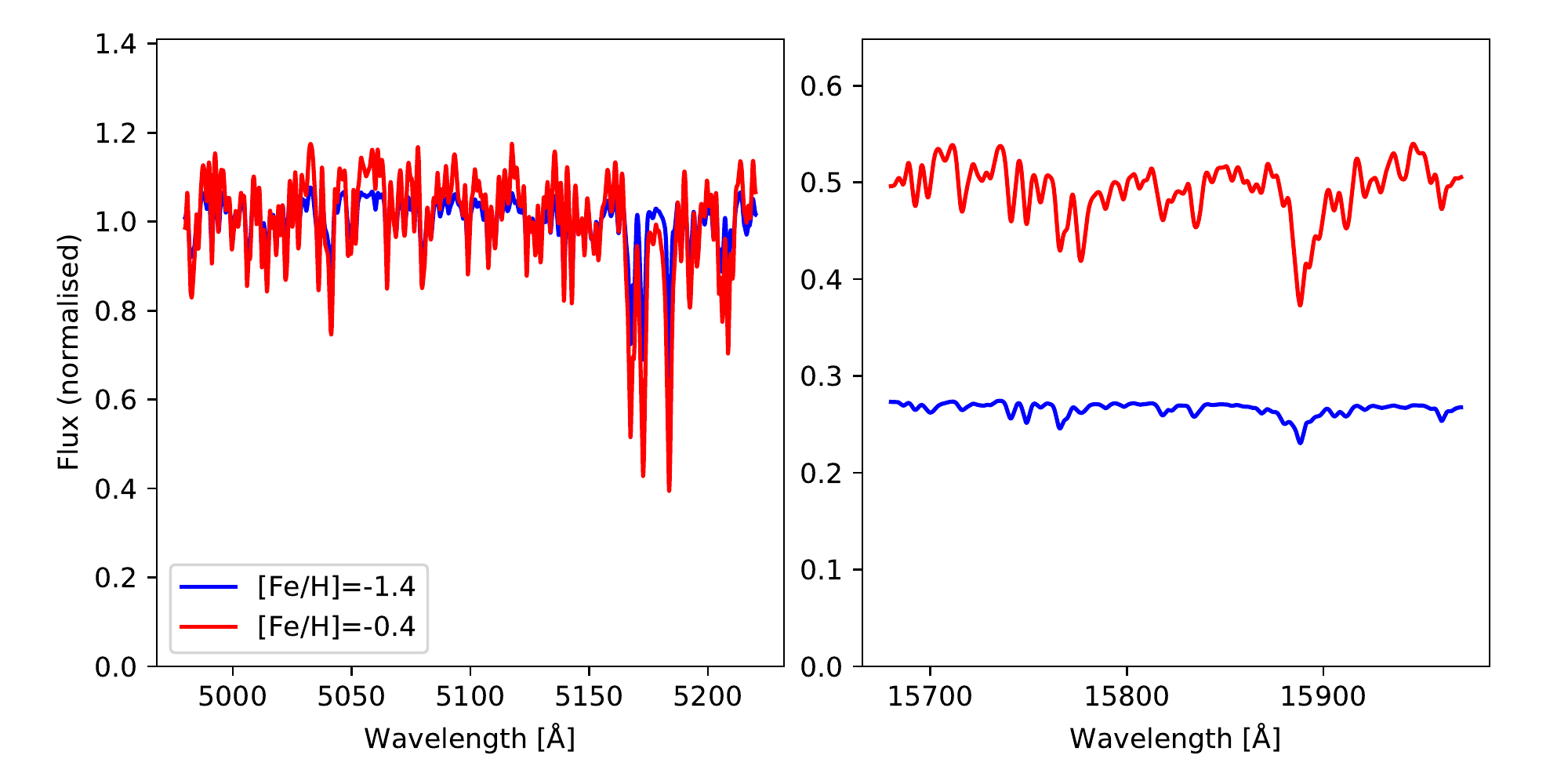}
      \caption{Model spectra for metal poor and metal rich populations with ages of 11 Gyr. The scaling is normalised at 5000--5200~\AA .
         \label{fig:p2pop}
         }
   \end{figure}

   \begin{figure}
   \centering
   \includegraphics[width=\columnwidth]{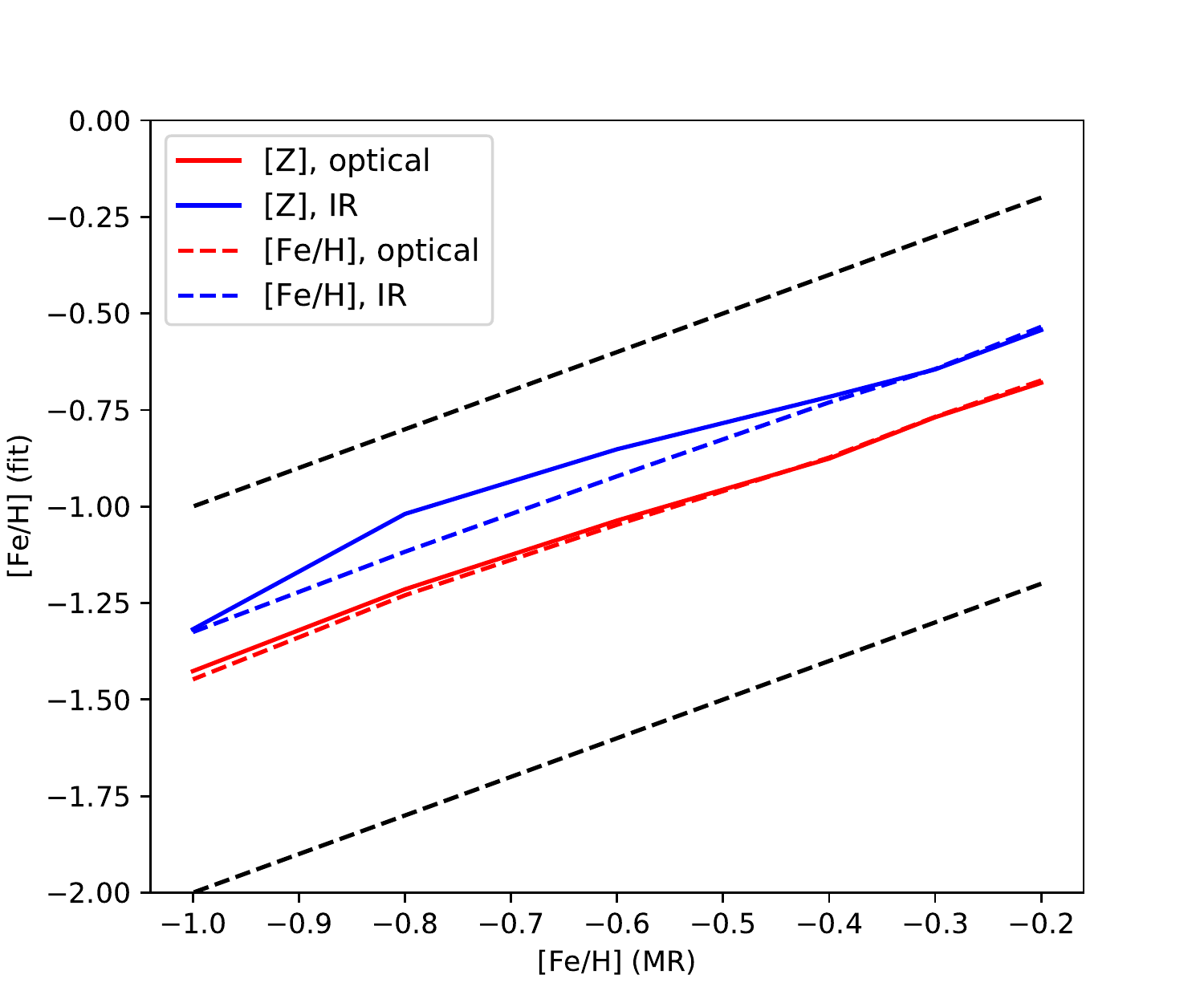}
      \caption{Metallicities derived from single-population fits to two-population models in the optical and infrared, plotted as a function of the input metallicity of the metal rich (MR) population. The solid coloured lines show fits where a global scaling was applied to all elements (labelled [Z]) whereas coloured dashed lines show fits where only Fe was varied (labelled [Fe/H]). The black dashed lines show the input metallicities.
         \label{fig:cmp2pop}
         }
   \end{figure}
   
      \begin{figure}
   \centering
   \includegraphics[width=\columnwidth]{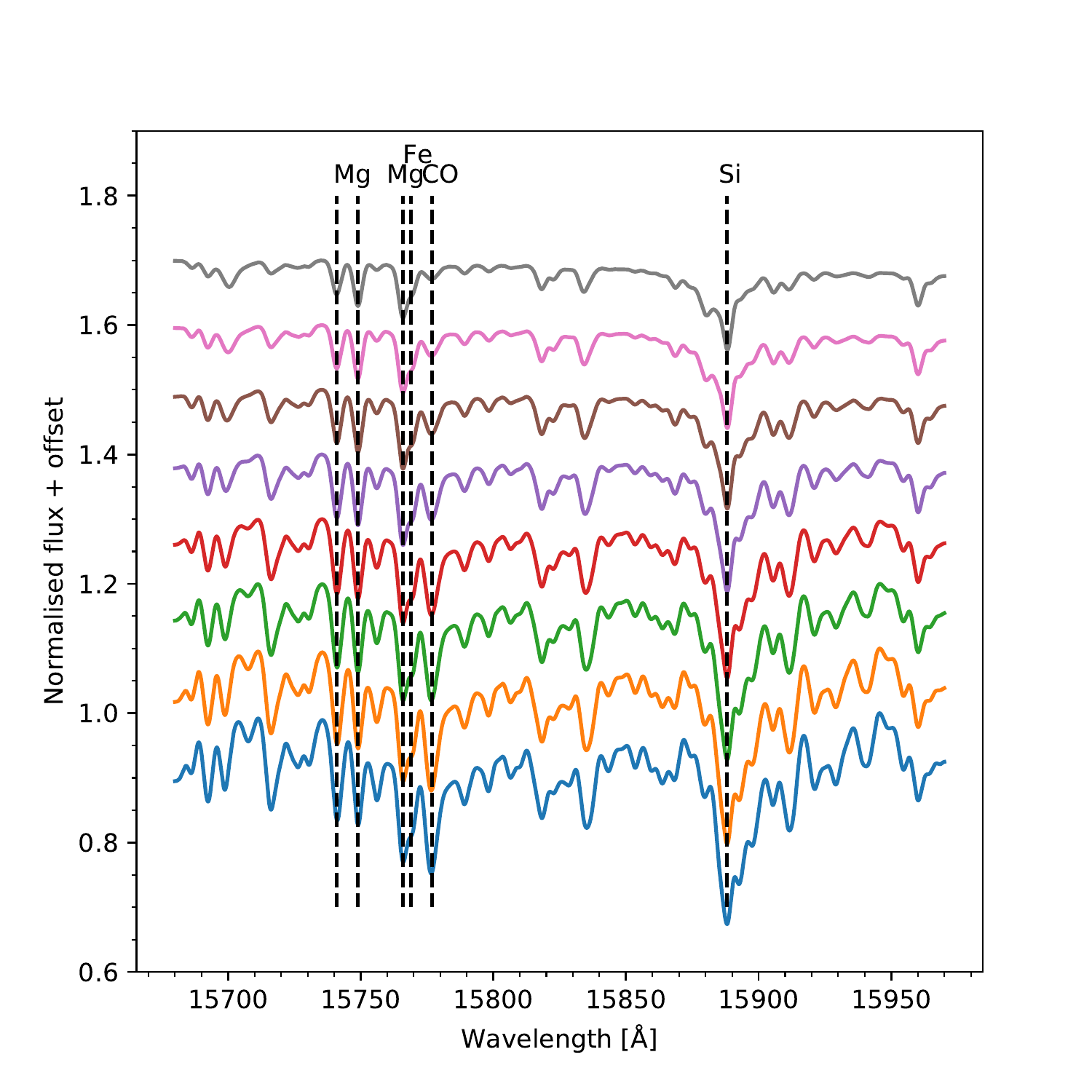}
      \caption{Model integrated-light infrared spectra for different metallicities, increasing from $\mathrm{[Fe/H]}=-0.2$ (bottom) to $\mathrm{[Fe/H]}=-1.6$ (top). A few strong features are labelled.
         \label{fig:irspec}
         }
   \end{figure}

   \begin{figure}
   \centering
   \includegraphics[width=\columnwidth]{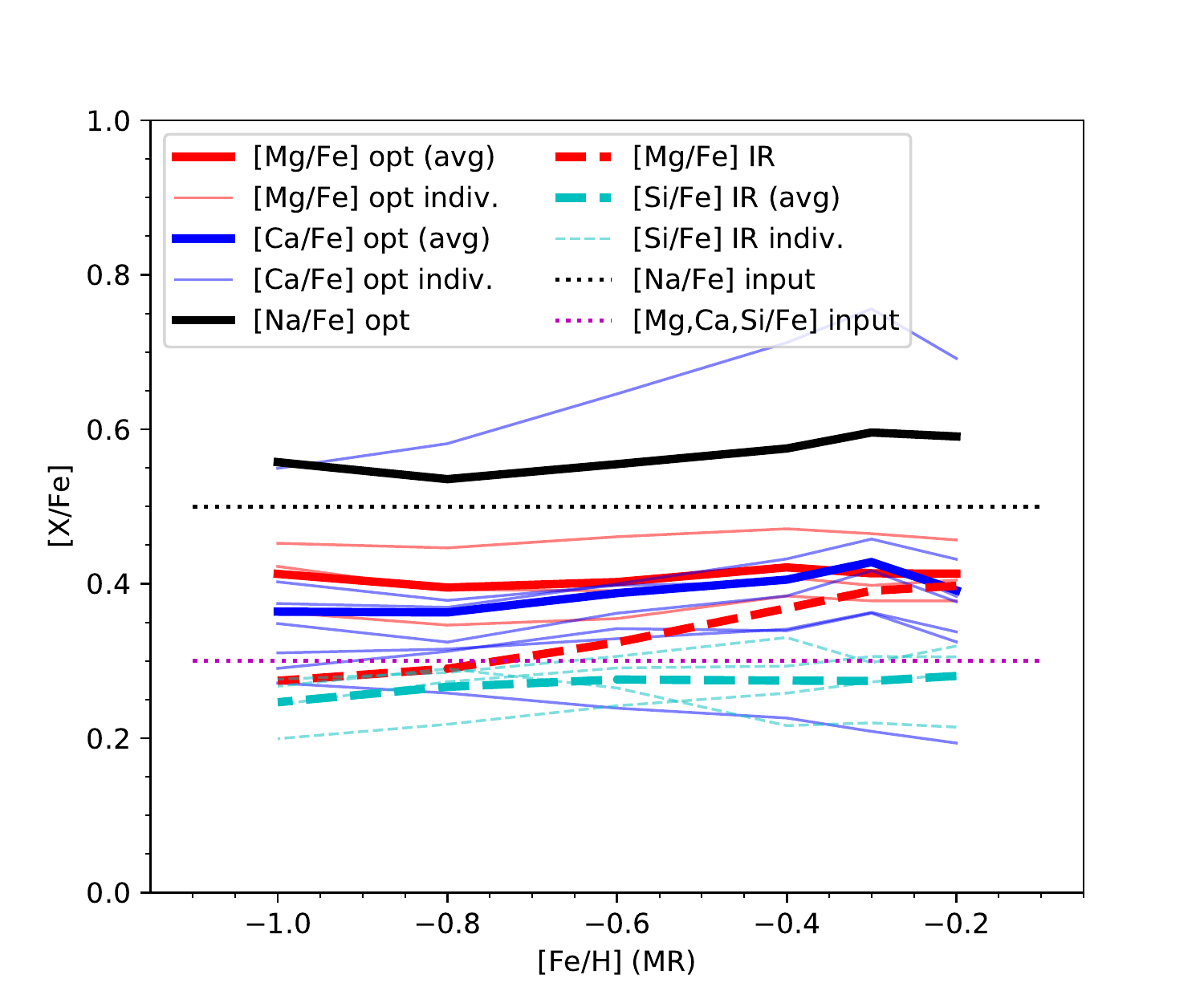}
      \caption{Abundances of individual elements derived from single-population fits to two-population models. Thick lines indicate average abundances and thin lines show measurements for individual bins.
      The input abundances were $\mathrm{[Mg/Fe]} = \mathrm{[Ca/Fe]} = \mathrm{[Si/Fe]} = +0.3$ and $\mathrm{[Na/Fe]} = +0.5$ for both populations, as indicated by the horizontal dotted lines.
         \label{fig:cmpxfe}
       }  
   \end{figure}

An important outcome of the analysis presented in the preceding sections is that we can obtain consistent results over a broad wavelength range, from the optical to the near-infrared. Our average integrated-light metallicity for G280 is slightly higher than that of 47~Tuc and almost identical to those found from optical spectra by \citet{Colucci2014}, and from near-infrared spectra by \citet{Sakari2016}. We also find similar abundance ratios for most individual elements as the previous studies. These results do not depend strongly on details on the HRD modelling, but we note that the older CH04 line list (used in our previous work on integrated-light abundance analysis) produces more discrepant results in the infrared, compared with our optical measurements and previous work. 

As noted in the introduction, several of the more massive Galactic GCs have substantial metallicity spreads. The best known example is $\omega$ Cen, although that cluster is more metal-poor than G280. The metallicity distribution in $\omega$ Cen peaks at $\mathrm{[Fe/H]}\simeq-1.7$, with some stars reaching metallicities as high as $\mathrm{[Fe/H]}\simeq-1$ \citep{Suntzeff1996,An2017} and the dispersion is about $\sigma_\mathrm{[Fe/H]}=0.3$ dex \citep{Willman2012}. For the massive globular cluster G1 in M31, \citet{Meylan2001} found a mean metallicity of $\mathrm{[Fe/H]}=-0.95$ and a metallicity dispersion of $\sigma_\mathrm{[Fe/H]}=0.4-0.5$ dex. While a few other Galactic GCs also have significant metallicity spreads, the dispersions are generally smaller than in $\omega$ Cen \citep{Carretta2009c,Willman2012}. With a $V$-band luminosity of $\log L_V = 6.24$ or $M_V=-10.78$ \citep{Strader2011}, G280 is both more massive and more luminous than $\omega$ Cen, and comparable to G1. In the following we consider the effects of a metallicity spread in G280 along the lines suggested by Fig.~\ref{fig:cmd} and previous analyses of the CMD (F08).

In general, the RGB becomes redder at higher metallicities, and the relative contributions from metal poor and metal rich stars to the integrated light are therefore dependent on wavelength. At longer wavelengths, the contribution of metal rich stars is expected to increase, and we may therefore expect that the average metallicity derived from an integrated-light measurement will be higher in the infrared, as compared to a measurement at optical wavelengths. Qualitatively, this is consistent with the difference between our NIRSPEC and HIRES measurements.
However, predicting the exact relation between the average integrated-light metallicity and the relative contributions of the various populations must account for the fact that the strengths of spectral features depend on metallicity in a non-linear way. 

To put this discussion on a more quantitative footing, we carried out a set of simulations in which we modelled the integrated-light spectra of clusters consisting of two populations, both with an age of 11 Gyr, with metallicities differing by 1 dex. We adopted an enhancement of the $\alpha$-elements of $+0.30$ dex and $\mathrm{[Na/Fe]}=+0.5$ dex. The HRD modelling was based on Dartmouth isochrones and for simplicity we ignored stellar evolutionary phases beyond the tip-RGB in these tests. The spectra of the metal poor and metal rich populations were normalised to each contribute 50\% of the flux when averaged over the interval 5000--5200 \AA. We then fitted these model spectra following our usual procedure. 
We note that the dispersion of a distribution consisting of two $\delta$-functions separated by 1 dex corresponds to $\sigma = 0.5$ dex. Hence, while the input metallicity distribution adopted here is clearly very artificial, the effects on the integrated light may be comparable (at least in a qualitative sense) to those of the largest metallicity spreads observed in actual GCs.

In Fig.~\ref{fig:p2pop} we show integrated-light spectra for two models with  $\mathrm{[Fe/H]}=-0.4$ and $\mathrm{[Fe/H]}=-1.4$. 
This figure illustrates how the relative contributions of the two populations to the flux depend on wavelength.
By design, the metal poor and metal rich populations each contribute 50\% of the flux in the optical, but at 1.6~$\mu$m the metal-rich population contributes almost twice as much to the flux as the metal-poor population (65\% versus 35\%, respectively).

Figure~\ref{fig:cmp2pop} shows the metallicities derived from fits to the optical and infrared integrated-light model spectra. 
The solid coloured lines show the metallicities obtained by allowing the global scaling of all abundances to vary, while the dashed coloured lines show the Fe abundance obtained by fitting specifically for $\mathrm{[Fe/H]}$ after fixing the global scaling to the value obtained from the global fits. We used the same input HRDs for the optical and infrared spectra, so that any offsets may be compared with the results for G280 in Tables~\ref{tab:abun-opt} and \ref{tab:abun-ir1}.
We see that the fits to the optical spectra generally return metallicities close to the average of the two populations, with very similar results for the global fits and the Fe-only fits. As expected, the fits to the infrared spectra generally return higher metallicities, and the offset tends to be more pronounced for the global fits than for the Fe-only fits. 
The offset between the optical and infrared fits depends on metallicity but reaches a maximum of about 0.20 dex for the global fits at intermediate metallicities. The maximum offset for the Fe-only fits is about 0.14 dex.
This difference is most likely caused by the non-linear behaviour of different features as a function of metallicity. As an example, Fig.~\ref{fig:irspec} shows infrared spectra for metallicities between $\mathrm{[Fe/H]}=-0.2$ and $\mathrm{[Fe/H]}=-1.6$. We note, for example, the change in the relative strengths of the atomic features (Fe, Mg) and the CO band-head at 15776~\AA .

For the most metal rich models in Fig.~\ref{fig:cmp2pop}, the average fitted Fe abundances are $\mathrm{[Fe/H]} = -0.673$ for the optical fits and $\mathrm{[Fe/H]} = -0.535$ in the infrared.  The difference of $\sim0.14$ dex is thus comparable to the $0.08\pm0.07$ dex offset between the Fe abundances obtained from the HIRES and NIRSPEC spectra of G280.
Given the random and systematic uncertainties on the measurements (as discussed in Sec.~\ref{sec:modasump}), it would be premature to claim that the spectroscopy presents strong evidence for a metallicity spread in G280, although it is certainly consistent with one.

Figure~\ref{fig:cmpxfe} shows the abundance ratios measured from our two-population model spectra as a function of metallicity (of the metal-rich component) for a selection of individual elements for the optical and infrared ranges. The mean abundance ratios, averaged over several features, are drawn with thick lines whereas the individual measurements are drawn with thin lines. We note that the individual measurements can vary by several tenths of a dex. It is clear that metallicity spreads can lead to significant systematic effects in integrated-light abundances, even if the abundance patterns themselves do not vary as a function of metallicity. For example, Fig.~\ref{fig:cmpxfe} suggests that the $\mathrm{[Na/Fe]}$ and $\mathrm{[Mg/Fe]}$ ratios may be systematically overestimated by $\sim0.1$ dex in G280 if the cluster has a substantial metallicity spread.

\section{Summary and conclusions}

We measured chemical abundances of the massive globular cluster G280 in M31 using optical and $H$-band integrated-light spectra. We also investigated how the presence of a metallicity spread within the cluster would affect the measured abundances.
Our main findings are as follows:

\begin{itemize}
\item When using the most recent line list from the Kurucz web site, we obtain consistent overall metallicities from the optical and $H$-band spectra ($\mathrm{[Fe/H]}=-0.68$ with rms$_w = 0.13$ dex, and $\mathrm{[Fe/H]}=-0.60$ with rms$_w=0.15$ dex, respectively). These measurements agree well with previously published metallicity determinations for G280.
\item The $\alpha$-elements are found to be enhanced by 0.3--0.4 dex. We also find sodium to be enhanced with $\mathrm{[Na/Fe]}=+0.57\pm0.05$ dex.
\item The results are relatively insensitive to the details of the modelling of the Hertzsprung-Russell diagram. When using the empirical colour-magnitude diagram for 47~Tuc or a MIST isochrone instead of the Dartmouth isochrone assumed in our standard analysis, the abundances measured from the optical spectrum change by less than 0.1 dex.  Larger changes (in particular for oxygen) are seen in the $H$-band, where the spectral features are more sensitive to differences in the properties of the coolest giants.
\item At the metallicity of G280, we find that a metallicity spread of $\sigma_\mathrm{[Fe/H]}\approx0.5$ dex within the cluster should lead to a small, but potentially detectable difference ($\sim0.14$ dex) between the optical and near-infrared metallicity estimates. While a small difference is indeed seen, it is entirely consistent with measurement errors.  
\item Our tests suggest that a similar metallicity spread would also cause the $\mathrm{[Na/Fe]}$  and the average $\mathrm{[Mg/Fe]}$ and $\mathrm{[Ca/Fe]}$ ratios to be overestimated by $\sim0.1$ dex. However, the sensitivity of specific features to abundance spreads can vary significantly. 
\end{itemize}

While the performance of our analysis technique had previously been tested on optical spectra, we have shown here that it also performs well in the $H$-band. This represents an important step towards refining this type of analysis for observations with future 30--40 m class telescopes, which will likely utilise the near-infrared more extensively than current facilities do. 

We draw attention to the fact that luminosity-weighted mean quantities, such as metallicities and abundance ratios, may respond differently to the presence of metallicity- and abundance spreads at different wavelengths and for different specific features. A corollary to this is that observations of multiple features may help constrain the presence of multiple populations with distinct metallicities and/or abundance patterns. However, additional work is required to identify the most suitable features and optimal wavelength ranges for such purposes, and to quantify observational requirements such as signal-to-noise ratios and spectral resolution.

\begin{acknowledgements}
JPB acknowledges support from NSF grant AST-1518294.
The data presented here were obtained at the W. M. Keck Observatory, which is operated as a scientific partnership among the California Institute of Technology, the University of California, and the National Aeronautics and Space Administration. The Observatory was made possible by the generous financial support of the W. M. Keck Foundation. The authors wish to recognise and acknowledge the very significant cultural role and reverence that the summit of Maunakea has always had within the indigenous Hawaiian community.  We are most fortunate to have the opportunity to conduct observations from this mountain.
We thank L.\ Origlia for helpful discussions about the modelling of near-IR spectra and the anonymous referee for a number of helpful comments. 
\end{acknowledgements}

\bibliographystyle{aa}
\bibliography{refs.bib}

\begin{appendix}
\section{Individual abundance measurements}
\longtab[1]{
\begin{longtable}{lcc}
\caption{\label{tab:ab-opt} Individual abundance measurements for HIRES data}\\
\hline\hline
Wavelength [\AA] & Value & Error \\ \hline
\endfirsthead
\caption{continued.}\\
\hline\hline
Wavelength [\AA] & Value & Error \\ \hline
\endhead
\hline
\endfoot
\mbox{[Fe/H]} \\
4024.2--4081.9 & $-0.763$ & 0.016 \\
4070.5--4128.8 & $-0.630$ & 0.040 \\
4117.8--4176.8 & $-0.933$ & 0.026 \\
4166.2--4225.9 & $-1.164$ & 0.025 \\
4215.8--4276.3 & $-0.744$ & 0.021 \\
4266.6--4327.8 & $-0.712$ & 0.015 \\
4318.6--4380.6 & $-0.593$ & 0.021 \\
4371.9--4434.6 & $-0.675$ & 0.010 \\
4426.6--4490.0 & $-0.672$ & 0.016 \\
4482.6--4546.9 & $-0.543$ & 0.016 \\
4540.0--4605.1 & $-1.024$ & 0.020 \\
4599.0--4664.9 & $-0.813$ & 0.026 \\
4659.5--4726.3 & $-0.784$ & 0.025 \\
4721.6--4789.3 & $-0.530$ & 0.020 \\
4785.4--4854.1 & $-0.573$ & 0.021 \\
4850.9--4920.5 & $-0.734$ & 0.011 \\
4918.2--4988.8 & $-0.582$ & 0.011 \\
4987.5--5059.1 & $-0.662$ & 0.012 \\
5058.8--5131.4 & $-0.481$ & 0.013 \\
5132.1--5205.7 & $-0.683$ & 0.013 \\
5207.5--5282.2 & $-0.664$ & 0.013 \\
5285.2--5361.1 & $-0.833$ & 0.015 \\
5365.3--5442.3 & $-0.633$ & 0.005 \\
5447.8--5526.0 & $-0.742$ & 0.021 \\
5532.9--5612.3 & $-0.725$ & 0.017 \\
5620.8--5701.4 & $-0.616$ & 0.016 \\
5711.4--5793.3 & $-0.433$ & 0.020 \\
5805.0--5888.2 & $-0.696$ & 0.025 \\
5901.8--5986.3 & $-0.563$ & 0.030 \\
\mbox{[Na/Fe]} \\
5677.0--5695.0 & $+0.565$ & 0.051 \\
\mbox{[Mg/Fe]} \\
4347.0--4357.0 & $+0.614$ & 0.066 \\
4565.0--4576.0 & $+0.225$ & 0.084 \\
4700.0--4707.0 & $+0.606$ & 0.030 \\
\mbox{[Ca/Fe]} \\
4222.0--4232.0 & $+0.045$ & 0.026 \\
4280.0--4320.0 & $+0.205$ & 0.051 \\
4420.0--4430.0 & $+0.526$ & 0.061 \\
4432.0--4460.0 & $+0.421$ & 0.025 \\
4575.0--4591.0 & $+0.186$ & 0.052 \\
4873.0--4883.0 & $+0.755$ & 0.088 \\
5259.0--5268.0 & $+0.874$ & 0.041 \\
5580.0--5608.0 & $+0.271$ & 0.035 \\
\mbox{[Sc/Fe]} \\
4290.0--4328.0 & $+0.164$ & 0.096 \\
4350.0--4380.0 & $+0.235$ & 0.111 \\
4383.0--4435.0 & $+0.362$ & 0.081 \\
4665.0--4675.0 & $+0.008$ & 0.151 \\
5026.0--5036.0 & $+0.474$ & 0.112 \\
5638.0--5690.0 & $+0.326$ & 0.026 \\
\mbox{[Ti/Fe]} \\
4292.0--4320.0 & $+0.425$ & 0.026 \\
4386.0--4420.0 & $+0.423$ & 0.030 \\
4440.0--4474.0 & $+0.345$ & 0.023 \\
4530.0--4542.0 & $+0.116$ & 0.033 \\
4545.0--4574.0 & $+0.515$ & 0.026 \\
4587.0--4593.0 & $+0.044$ & 0.105 \\
4650.0--4660.0 & $+0.465$ & 0.055 \\
4665.0--4715.0 & $+0.554$ & 0.040 \\
4750.0--4784.0 & $+0.115$ & 0.043 \\
4790.0--4850.0 & $+0.486$ & 0.038 \\
4993.0--5045.0 & $+0.306$ & 0.019 \\
5152.5--5160.0 & $+0.775$ & 0.299 \\
\mbox{[Cr/Fe]} \\
4250.0--4270.0 & $+0.045$ & 0.036 \\
4272.0--4292.0 & $-0.175$ & 0.039 \\
4350.0--4375.0 & $-0.713$ & 0.081 \\
4375.0--4400.0 & $+0.006$ & 0.098 \\
4520.0--4547.0 & $-0.085$ & 0.037 \\
4547.0--4605.0 & $-0.424$ & 0.030 \\
4605.0--4660.0 & $-0.076$ & 0.030 \\
5235.0--5282.0 & $+0.328$ & 0.020 \\
5285.0--5330.0 & $-0.076$ & 0.024 \\
5342.0--5351.0 & $+0.165$ & 0.045 \\
5407.0--5413.0 & $+0.478$ & 0.041 \\
\mbox{[Mn/Fe]} \\
4750.0--4780.0 & $-0.080$ & 0.041 \\
\mbox{[Ni/Fe]} \\
4600.0--4610.0 & \multicolumn{2}{c}{$\ldots$} \\
4644.0--4654.0 & $+0.645$ & 0.066 \\
4681.0--4691.0 & $-0.543$ & 0.138 \\
4709.0--4719.0 & $+0.510$ & 0.051 \\
4824.0--4835.0 & $-0.054$ & 0.121 \\
4899.0--4909.0 & $+0.216$ & 0.090 \\
4931.0--4942.0 & $+0.536$ & 0.081 \\
4975.0--4985.0 & $+0.304$ & 0.060 \\
5075.0--5089.0 & $-0.046$ & 0.042 \\
5098.0--5108.0 & $+0.295$ & 0.066 \\
5141.0--5151.0 & $+0.515$ & 0.085 \\
5472.0--5482.0 & $+0.016$ & 0.030 \\
\mbox{[Ba/Fe]} \\
4551.0--4560.0 & $-0.155$ & 0.056 \\
4929.0--4939.0 & $-0.184$ & 0.106 \\
5849.0--5859.0 & $-0.455$ & 0.111 \\
\end{longtable}
}
\longtab[2]{
\begin{longtable}{lccc}
\caption{\label{tab:ab-ir} Individual abundance measurements for NIRSPEC data. The last column gives the fraction of the total number of pixels available for the fit after masking out telluric absorption.}\\
\hline\hline
Wavelength [\AA] & Value & Error & F(used) \\ \hline
\endfirsthead
\caption{continued.}\\
\hline\hline
Wavelength [\AA] & Value & Error & F(used) \\ \hline
\endhead
\hline
\endfoot
\mbox{[Fe/H]} \\
15395.0--15623.2 & $-0.775$ & 0.044 & 1.000 \\
15713.5--15946.2 & $-0.635$ & 0.021 & 0.875 \\
16044.6--16283.0 & $-0.335$ & 0.036 & 0.921 \\
16389.8--16634.8 & $-0.714$ & 0.046 & 0.958 \\
16746.3--17002.7 & $-0.465$ & 0.074 & 0.976 \\
17129.9--17385.3 & $-1.105$ & 0.226 & 0.936 \\
\mbox{[C/Fe]} \\
15395.0--15623.2 & $+0.124$ & 0.009 & 1.000 \\
15713.5--15946.2 & $-0.116$ & 0.062 & 0.875 \\
16044.6--16283.0 & $+0.042$ & 0.026 & 0.921 \\
16389.8--16634.8 & $-0.436$ & 0.054 & 0.958 \\
16746.3--17002.7 & $-0.116$ & 0.058 & 0.976 \\
17129.9--17385.3 & $+0.003$ & 0.077 & 0.936 \\
\mbox{[O/Fe]} \\
15395.0--15623.2 & $+0.662$ & 0.028 & 1.000 \\
15713.5--15946.2 & $+0.292$ & 0.042 & 0.875 \\
\mbox{[Mg/Fe]} \\
15730.0--15770.0 & $+0.163$ & 0.106 & 0.754 \\
\mbox{[Si/Fe]} \\
15850.0--15920.0 & $+0.342$ & 0.036 & 1.000 \\
16070.0--16250.0 & $+0.414$ & 0.060 & 0.912 \\
16420.0--16470.0 & $+0.382$ & 0.339 & 0.947 \\
17200.0--17350.0 & $+0.163$ & 0.161 & 0.942 \\
\end{longtable}
}

\longtab[3]{
\begin{longtable}{lccccc}
\caption{\label{tab:ab-irn} Individual abundance measurements for NIRSPEC data, fitting multiple elements simultaneously.}\\
\hline\hline
Wavelength [\AA] & $\mathrm{[Fe/H]}$ & $\mathrm{[C/Fe]}$ & $\mathrm{[N/Fe]}$ & $\mathrm{[O/Fe]}$ & $\mathrm{[Si/Fe]}$ \\ \hline
\endfirsthead
\caption{continued.}\\
\hline\hline
Wavelength [\AA] & Value & Error & \\ \hline
\endhead
\hline
\endfoot
15395.0-15623.2   & $-0.540$   & $-0.090$   & $+1.450$   & $+0.393$   & $\ldots$  \\
15713.5-15946.2   & $-0.714$   & $-0.279$   & $\ldots$   & $+0.426$   & $+0.431$  \\
16044.6-16283.0   & $-0.461$   & $-0.109$   & $\ldots$   & $+0.168$   & $+0.388$  \\
16389.8-16634.8   & $-0.643$   & $-0.370$   & $\ldots$   & $+0.542$   & $-0.143$  \\
16746.3-17002.7   & $-0.541$   & $-0.256$   & $\ldots$   & $+0.145$   & $\ldots$  \\
17129.9-17385.3   & $-0.983$   & $+0.433$   & $\ldots$   & $+0.982$   & $+0.660$  \\
\end{longtable}
}

\end{appendix}

\end{document}